\documentclass{jfm}
\usepackage{graphicx,color}
\usepackage{comment}
\usepackage{bm}
\usepackage{subcaption}
\usepackage{mathtools}
\usepackage{amsmath,amssymb}
\newcommand{\GK} [1] {{\color{black}#1}}

\begin{document}

\newtheorem{lemma}{Lemma}
\newtheorem{corollary}{Corollary}

\shorttitle{Ultimate heat transfer in `wall-bounded' convective turbulence} %for header on odd pages
\shortauthor{K. Kawano, S. Motoki, M. Shimizu and G. Kawahara} %for header on even pages

\title{Ultimate heat transfer in `wall-bounded' convective turbulence}
\author{Koki Kawano\aff{1}, Shingo Motoki\aff{1}, Masaki Shimizu\aff{1}\\ \and Genta Kawahara\aff{1}\corresp{\email{kawahara@me.es.osaka-u.ac.jp}}}
\affiliation{\aff{1}Graduate School of Engineering Science, Osaka University, 1-3 Machikaneyama, Toyonaka, Osaka 560-8531, Japan}

\maketitle

\begin{abstract}
Direct numerical simulations have been performed for turbulent thermal convection between horizontal no-slip, permeable walls with a distance $H$ and a constant temperature difference $\Delta T$ \GK{at the Rayleigh number $Ra=3\times 10^3$--$10^{10}$}.
On the no-slip wall surfaces $z=0$, $H$ the wall-normal (vertical) transpiration velocity is assumed to be proportional to the local pressure fluctuation, i.e. $w=-\beta p'/\rho$, $+\beta p'/\rho$ (Jim\'enez {\it et al.}, {\it J. Fluid Mech.}, vol. 442, 2001, pp. 89--117), and the property of the permeable wall is \GK{given} by the permeability parameter \GK{$\beta U$} \GK{normalised with} the buoyancy-induced terminal velocity \GK{$U=(g\alpha\Delta TH)^{1/2}$}, where $\rho$, $g$ and $\alpha$ are \GK{mass density, acceleration due to gravity and volumetric thermal expansivity}, respectively.
A zero net mass flux through the wall is instantaneously ensured, and thermal convection is driven \GK{only} by buoyancy without \GK{any} additional energy \GK{inputs}.
\GK{The critical transition of heat transfer in convective turbulence has been found} between \GK{the two $Ra$ regimes} for fixed \GK{$\beta U=3$} and \GK{fixed} Prandtl number $Pr=1$.
In the subcritical regime \GK{at lower $Ra$} the Nusselt number $Nu$ scales with $Ra$ as $Nu\sim Ra^{1/3}$, as commonly observed in turbulent Rayleigh--B\'enard convection.
In the supercritical regime \GK{at higher $Ra$}, on the other hand, the ultimate scaling $Nu\sim Ra^{1/2}$ is achieved,
\GK{meaning that the wall-to-wall heat flux scales with $U\Delta T$ independent of the thermal diffusivity,
although the heat transfer on the wall is dominated by thermal conduction.}
In \GK{an} impermeable case \GK{($\beta U=0$) as well as} even in the subcritical \GK{permeable} case the vertical velocity fluctuation is weak near the wall, and \GK{it scales with $Ra^{-1/6}U$ corresponding to the velocity scale of near-wall small-scale thermal plumes.}
In the supercritical \GK{permeable} case, contrastingly, \GK{large-scale} motion is induced \GK{by buoyancy even} in the vicinity of the wall, \GK{leading to significant transpiration velocity of the order of $U$.}
\GK{The ultimate heat transfer is attributed to this large-scale significant fluid motion.}
\GK{In such `wall-bounded' convective turbulence, a thermal conduction layer still exists on the wall, but there is no near-wall layer of large change in the vertical velocity, suggesting that the effect of the viscosity is negligible even in the near-wall region.}
The balance between the dominant advection and buoyancy terms in the vertical Boussinesq equation gives us the velocity scale of \GK{$O(U)$ in the whole region, so that
the total energy budget equation implies the Taylor's dissipation law \GK{$\epsilon\sim U^3/H$ and the ultimate scaling $Nu\sim Ra^{1/2}$}.}
\end{abstract}

\section{Introduction}
The flow driven by \GK{buoyancy} is called thermal convection, and it plays an important role in \GK{a wide variety of phenomena of} geophysics, astrophysics and engineering applications.
One of the \GK{canonical configurations of} thermal convection is \GK{the} Rayleigh--B\'enard convection (RBC) \GK{observed in a horizontal fluid layer heated from below and cooled from above}.
In RBC, buoyancy forcing is characterised \GK{in terms of} the Rayleigh number $Ra$ and the flow \GK{becomes turbulent eventually} as $Ra$ increases. 

It is known that the Nusselt number $Nu$ (dimensionless vertical heat flux) \GK{exhibits the power law of $Ra$, $Nu\sim Ra^\gamma$, for a certain value of $\gamma$ in the turbulent state of RBC}.
For more than half a century, various predictions have been made to clarify the scaling exponent $\gamma$.
\citet{priestley54} derived $\gamma=1/3$ from similarity \GK{argument}, and \citet{malkus54} also led to $\gamma=1/3$ based on the assumption that heat transfer is determined by \GK{the marginal instability of a} thermal boundary layer. 
\citet{kraichnan62} predicted $\gamma=1/2$ with a logarithmic correction, $Nu\sim Pr^{1/2}Ra^{1/2}(\ln{Ra})^{-3/2}$, as a scaling in a high-$Ra$ asymptotic state with turbulent boundary layers.
The scaling $Nu\sim Pr^{1/2}Ra^{1/2}$ is currently known as the ultimate scaling.
It has been derived as a rigorous upper bound on the heat transfer in RBC by applying variational methods to the Boussinesq equations \citep{doering92,doering96,plasting03}, and \GK{has recently been} obtained as a maximal heat transfer scaling between two parallel plates \citep{motoki18}.
The ultimate scaling \GK{relates} to the Taylor's energy dissipation law \GK{of} high-Reynolds-number turbulence \GK{via the rigorous energy budget equation of thermal convection}.
In the ultimate \GK{heat transfer} the energy dissipation and the scalar dissipation (corresponding to the vertical heat flux) are independent of the kinematic viscosity \GK{or} the thermal diffusivity.

Recently, \citet{GL00,GL02,GL11} have proposed a unifying scaling theory in RBC, and \GK{its} validity has been demonstrated by a lot of experimental and numerical studies \citep[\GK{see}][]{ahlers09rev}.
\GK{Their} theory is based on \GK{the energy budget equation relating the energy and scalar dissipation rates and on the decomposition of the flow field into a boundary layer and a bulk region}.
The theory gives different scaling laws depending on whether the \GK{total} energy and scalar dissipation rates are \GK{dominated by the bulk or the boundary layer}.
At \GK{a} high-$Ra$ regime, \GK{in which} the contribution from the bulk is dominant, the scaling $Nu\sim Ra^{1/3}$ is \GK{given if} the thermal boundary layer is thinner than the velocity boundary layer, \GK{but} the ultimate scaling $Nu\sim Pr^{1/2}Ra^{1/2}$ is \GK{anticipated if the thermal boundary layer is thicker than the velocity boundary layer}.

The question of whether or not the ultimate scaling \GK{can be achieved} has \GK{long} attracted a great deal of attention, and much effort has been spent on both experimental and numerical studies in the past few decades.
However, $Nu\sim Pr^{1/2}Ra^{1/2}$ has not been found as yet in conventional RBC, \GK{and what has been observed experimentally and numerically at high $Ra$ is the scaling} $Nu\sim Ra^{1/3}$ \citep[\GK{see}][]{ahlers09rev, chilla12rev}.
%According to the Kraichnan's prediction and the \GK{\citet{GL00,GL02,GL11}'s} theory, \GK{a logarithmic mean temperature profile should appear in} the turbulent thermal boundary layer in the ultimate state.
%\GK{In their experiments and numerical simulations for RBC, recently}, \cite{ahlers12log,ahlers14log} have \GK{found} that the temperature \GK{averaged} on GK{the circle of} a specific radius in a cylindrical vessel exhibits a logarithmic \GK{vertical} profile.
%\cite{van15} have also reported the logarithmic mean temperature in the \GK{horizontal positions} where thermal plumes are \GK{ejected} in two-dimensional RBC.
%In RBC, however, \GK{the} clear logarithmic temperature profile, as seen in turbulent \GK{wall-bounded} shear flows, has not been observed even at high $Ra$.

\GK{It is known that the ultimate scaling $Nu\sim Pr^{1/2}Ra^{1/2}$ can be \GK{observed} in turbulent thermal convection without horizontal bounding walls on which thermal and velocity boundary layers should have appeared.}
Such \GK{wall-less} thermal convection \GK{was numerically examined} in a triply-periodic domain with a constant temperature gradient in the vertical direction \citep{calzavarini05}, and \GK{was experimentally investigated} in a vertical tube connecting high- and low-temperature chambers \citep{pawar16}.
The ultimate scaling has \GK{also} been reported for the thermal convection in a cylindrical container \GK{radiatively heated from below, instead of conventional RBC heating} \citep{lepot18,bouillaut19}.
In the radiatively-driven convection $Nu\sim Ra^{1/3}$ \GK{has been} observed when the thickness of heating layer is thin, and the scaling \GK{has been found to change} to $Nu\sim Pr^{1/2}Ra^{1/2}$ with increasing the thickness.

\GK{In case of conventional RBC heating, it has been found that surface roughness on horizontal walls transiently yields the ultimate scaling $Nu\sim Pr^{1/2}Ra^{1/2}$ in the limited range of $Ra$} where the thermal \GK{conduction} layer thickness is comparable to the size of roughness \GK{elements} \citep{zhu17,zhu19,tummers19}.
\GK{This transient} scaling would not imply the transition to the asymptotic ultimate scaling, because a further increase in $Ra$ leads to \GK{saturation down to} the usual scaling $Nu\sim Ra^{1/3}$.
It is still an open question whether or not the ultimate \GK{heat transfer} can be achieved by introducing \GK{an ingenious contrivance, such as wall roughness and so on, into wall-bounded RBC heated conventionally}. 

\GK{In this study, we introduce wall permeability into RBC.
\citet{jimenez01} have investigated turbulent momentum transfer in numerically simulated porous channel flow to find out that the wall permeability significantly enhances momentum transfer.
In their simulation the fluid crosses the porous wall surface with a wall-normal velocity proportional to pressure fluctuations.
This boundary condition mimics the behaviour of a zero-pressure-gradient boundary layer over a Darcy-type porous wall \citep[pp. 223--224]{batchelor67} with a constant-pressure plenum chamber underneath.}
\GK{We perform direct numerical simulations (DNS) for convective turbulence between horizontal no-slip, mass-neutral permeable walls with a constant temperature difference for fixed Prandtl number $Pr=1$ by using %\citet['s][]{jimenez01} 
Jim\'enez {\it et al.}'s (2001) boundary condition on a permeable wall.}
We report that the wall permeability brings about the ultimate \GK{heat transfer} $Nu\sim Ra^{1/2}$ at a high Rayleigh number in spite of the presence of \GK{a} thermal \GK{conduction} layer \GK{on the walls}.
\GK{We inspect scaling laws and turbulence structure in thermal convection between the permeable walls as well as impermeable walls to discuss why the ultimate heat transfer can be achieved by the introduction of permeable walls.}

This paper is organised as follows.
The numerical procedure to solve the Boussinesq equations with the no-slip, permeable boundary conditions is presented in \S \ref{sec:dns}, and it is confirmed that there \GK{are no additional energy inputs} except \GK{for buoyancy power} in \S \ref{sec:budget}.
\GK{Scaling properties and turbulence structure} in thermal convection between permeable and impermeable walls are presented in \S \ref{sec:result}, and the physical interpretation of the scaling laws is provided in \S \ref{sec:mechanism}.
The summary and \GK{outlook} are given in \S \ref{sec:summary}.
The Prandtl number dependence of the scaling of $Nu$ with $Ra$ is \GK{briefly} shown in appendix \ref{sec:prandtl}, where it is demonstrated that the ultimate scaling can also be observed for the Prandtl number $Pr=7$.

\section{Direct numerical simulation}\label{sec:dns}

We conduct \GK{DNS} for turbulent thermal convection between horizontal plates with a distance $H$ and a constant temperature difference $\Delta T$.
The \GK{Oberbeck--Boussinesq} approximation is employed, wherein density variations are taken into account only in the buoyancy term.
The two horizontal and vertical directions are respectively denoted by $x$, $y$ and $z$ (or $x_{1}$, $x_{2}$ and $x_{3}$).
The corresponding components of the velocity $\bm{u}(\bm{x},t)$ are given by $u$, $v$ and $w$ (or $u_{1}$, $u_{2}$ and $u_{3}$) in the horizontal and vertical directions, respectively.

The governing equations are the Boussinesq equations
\begin{eqnarray}
\label{renzoku}
&\displaystyle
\nabla \cdot \bm{u}=0,&\\
\label{ns}
&\displaystyle
\frac{\partial \bm{u}}{\partial t}+(\bm{u} \cdot \nabla)\bm{u}=-\frac{1}{\rho} \nabla p +\nu \nabla^2 \bm{u} + g \alpha T\bm{e}_{z},&\\
\label{iryu}
&\displaystyle
\frac{\partial T}{\partial t} + (\bm{u} \cdot \nabla)T= \kappa \nabla^2 T,&
\end{eqnarray}
where $p(\bm{x},t)$ is the pressure, $T(\bm{x},t)$ is the temperature, and $\rho$, $\nu$, $g$, $\alpha$ and $\kappa$ are mass density, kinematic viscosity, acceleration due to gravity, a volumetric expansion coefficient and thermal diffusivity, respectively.
$\bm{e}_{z}$ is a unit vector in the vertical direction.
The velocity and temperature fields are supposed to be periodic in the horizontal ($x$- and $y$-) directions, and the periods \GK{in the $x$- and $y$-directions are taken to be} $L$.

\GK{We suppose that the two horizontal walls are composed of porous media with constant-pressure plenum chambers underneath and overhead.
The lower (or upper) wall and the associated plenum chamber are heated from below (or cooled from above).
On the permeable wall surface the vertical velocity $w$ is assumed to be proportional to the local pressure fluctuation $p'$ \citep{jimenez01}.}
\GK{The boundary conditions imposed on the walls are}
\begin{eqnarray}
\label{eq_bc_uv}
&\displaystyle
u(z=0)=u(z=H)=0, \;\; v(z=0)=v(z=H)=0,&\\
\label{eq_darcy}
&\displaystyle
w(z=0)=-\beta\frac{\GK{p'}}{\rho}, \;\; w(z=H)=\beta\frac{\GK{p'}}{\rho},&\\
&\displaystyle
T(z=0)=\Delta T, \;\; T(z=H)=0,&
\end{eqnarray}
where $\beta$ ($\ge0$) represents the \GK{property of permeability}, and the impermeability conditions $w(z=0,H)=0$ are recovered for $\beta=0$, while $\beta \to \infty$ implies zero pressure fluctuations \GK{and an unconstrained vertical velocity.
The flow situation observed in the thermal convection without horizontal walls \citep{calzavarini05}
is intuitively similar to this limit, although not identical.}
\GK{Note that a zero net mass flux through the permeable wall is instantaneously ensured because the transpiration velocity is proportional to the pressure {\it fluctuation} with zero mean.
We anticipate the no-slip and permeable conditions (\ref{eq_bc_uv}) and (\ref{eq_darcy}) on a wall with a large number of wall-normal through holes.} 
The proportionality coefficient $\beta$ has the \GK{dimension} of an inverse velocity, and \GK{thus $\beta U$ represents a dimensionless parameter determining the property of permeable walls if the buoyancy-induced terminal velocity $U=(g \alpha \Delta T H)^{1/2}$ is a proper velocity scale.
If the proper velocity scale (say, $U_w$) is smaller than $U$ as in the subcritical permeable case discussed later (see (\ref{eq:w_BL}) in \S\ref{sec:mechanism}), then the permeable condition $\beta U=\mbox{const}$. ($=\beta' U_w$) to be employed here implies a more permeable wall of larger $\beta'$ ($=(U/U_w)\beta$).}
\GK{Thermal} convection between permeable walls is characterised \GK{in terms of the Rayleigh number $Ra$, the Prandtl number $Pr$ and the permeability $\beta U$, where}
\begin{equation}
Ra=\frac{g \alpha \Delta T H^3}{\nu \kappa}, \;\; \quad Pr=\frac{\nu}{\kappa}. %, \;\; \Gamma=\frac{L}{H}.
\label{eq_rapras}
\end{equation}

The vertical heat flux from the bottom to the top wall is quantified by the Nusselt number $Nu$ \GK{written as}
\begin{equation}
Nu\GK{\equiv}-\frac{H}{\Delta T} \frac{{\rm d} \langle T \rangle_{xyt}}{ {\rm d} z} \bigg |_{z=0 }\GK{\equiv}-\frac{H}{\Delta T} \frac{{\rm d} \langle T \rangle_{xyt}}{ {\rm d} z} \bigg |_{z=H}
\GK{=1+\frac{H}{\kappa \Delta T} \langle wT \rangle_{xyzt}},
\label{eq_nu}
\end{equation}
where $\langle \cdot \rangle_{xyt}$ represents the average over the two horizontal directions and time, and $\langle \cdot \rangle_{xyzt}$ is the volume and time average.
The \GK{rightmost} equality is given by the \GK{volume and time} average of the \GK{energy} equation (\ref{iryu}).
\GK{Let us note that since the walls are isothermal in the permeable and impermeable cases, the temperature fluctuation and so the convective heat flux $\langle Tw \rangle_{xyt}$ are null on the walls ($z=0$, $H$) at any cases.
In the near-wall region, therefore, the conduction heat transfer dominates the convective one even in the permeable case.}

The Boussinesq equations (\ref{renzoku})--(\ref{iryu}) are discretised by employing a spectral Galerkin method based on the Fourier series expansion in the periodic horizontal directions and \GK{the Chebyshev polynomial expansion} in the vertical direction.
The nonlinear terms are evaluated using a spectral collocation method.
Aliasing errors are removed with the aid of the $2/3$ rule for the Fourier transform and the $1/2$ rule for the Chebyshev transform.
Time advancement is performed with the third-order Runge--Kutta scheme (\GK{or} the second-order Adams--Bashforth scheme) for the nonlinear and buoyancy terms and the implicit Euler scheme (\GK{or} the Crank--Nicolson scheme) for the diffusion terms in the permeable (\GK{or} impermeable) case.
The numerical procedure developed by \cite{jimenez01} is applied to satisfy the permeable boundary conditions.

In this paper, we \GK{shall} present the results \GK{obtained from DNS for thermal convection in the impermeable case $\beta U=0$ at $Ra=10^6-10^{11}$ and in the permeable case $\beta U=3$ at $Ra=3\times 10^3-10^{10}$} for \GK{fixed Prandtl number} $Pr=1$ and \GK{for fixed horizontal period $L/H=1$}.
The dependence on the Prandtl number is shown in appendix \ref{sec:prandtl}.
\GK{We have examined the dependence of heat transfer on the horizontal period in the range of $1\leq L/H\leq 4$ to confirm that the ultimate scaling $Nu\sim Ra^{1/2}$ to be shown in \S\ref{sec:result} can also be achieved for smaller $\beta U$ in a wider periodic box of larger $L/H$.}

\section{Energy budget}\label{sec:budget}

In this section, we \GK{discuss the total energy budget} in thermal convection between \GK{no-slip}, permeable walls.
By taking the \GK{volume} and time average of \GK{an} inner product of the Navier--Stokes equation (\ref{ns}) with the velocity ${\bm u}$ and \GK{taking account of the boundary conditions (\ref{eq_bc_uv}) and (\ref{eq_darcy})}, we obtain
\GK{
\begin{eqnarray}
\label{total energy budget}
\displaystyle
g \alpha  \langle w T \rangle_{xyzt}
=\epsilon
+\frac{1}{\beta H}\left( \left. {\left< w^{2} \right>}_{xyt} \right|_{z=0}+\left. {\left< w^{2} \right>}_{xyt}\right|_{z=H} \right)
+\frac{1}{2 H}{\left[ {\left< w^3 \right>}_{xyt} \right]}^{z=H}_{z=0}, \nonumber \\
\end{eqnarray}
where
\begin{equation}
\label{dissipation}
\epsilon=\frac{\nu}{2}\left\langle\left(\frac{\partial u_i}{\partial x_j}+\frac{\partial u_j}{\partial x_i}\right)^2\right\rangle_{xyzt}
\end{equation}
is a total energy dissipation rate per unit mass.}
\GK{The left-hand side of (\ref{total energy budget}) represents buoyancy power (energy input), while the second and the third terms on the right-hand side denote pressure power on the permeable walls and outflow kinetic energy across the permeable walls, respectively.
The pressure power on the permeable walls is strictly greater than zero, so that it is always an energy sink.
Although its sign cannot be specified rigorously, we have confirmed numerically that the outflow kinetic energy across the permeable walls is also positive in the present DNS, implying that the kinetic energy flows out of the system across the permeable walls.
It turns out that as in the impermeable case, thermal convection between the permeable walls is sustained only by the buoyancy power without any additional energy inputs.
It has also been found numerically that the pressure power is comparable with the energy dissipation whereas the outflow kinetic energy is much less than the dissipation.
The energy to be lost in the system via the permeable walls could be considered to be supplied to the other system, i.e. the flow in porous media, to eventually dissipate therein.

\begin{figure}
\centering
\begin{minipage}{0.8\linewidth}
\includegraphics[clip,width=\linewidth]{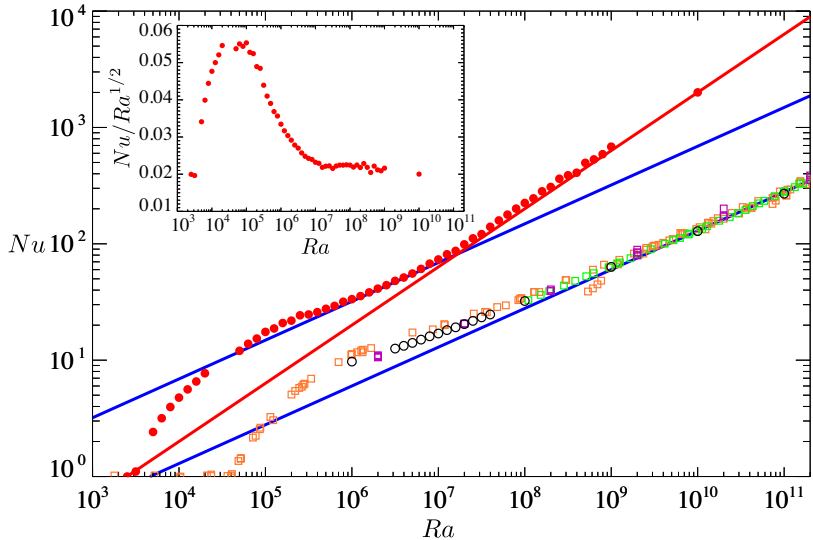}
\end{minipage}	
\caption{The Nusselt number $Nu$ as a function of the Rayleigh number $Ra$.
The open black and filled red circles respectively represent the present DNS data in the impermeable case $\beta U=0$ and permeable case $\beta U=3$ for the Prandtl number $Pr=1$.
The orange and green squares denote the experimental data in a cylindrical cell, taken from \cite{chavanne01} ($Pr \geq 0.7$) and \citeauthor{niemela06} ($Pr \geq 0.69$), respectively.
The purple squares stand for DNS data in a cylindrical cell, taken from \citeauthor{stevens10} ($Pr=0.7$).
The red and blue lines indicate the ultimate scaling $Nu\sim Ra^{1/2}$ and the ordinary scaling $Nu\sim Ra^{1/3}$, respectively.
The inset shows $Nu$ compensated by $Ra^{1/2}$ in the permeable case.
\label{fig:nu_ra}}
\end{figure}

The rightmost equality of (\ref{eq_nu}) yields the relation among the buoyancy power, the Prandtl number, the Rayleigh number and the Nusselt number given by
\begin{equation}
\label{total heat flux}
PrRa(Nu-1)=\frac{g \alpha  \langle w T \rangle_{xyzt} }{ {\kappa^3/H^4} }.
\end{equation}
Substituting (\ref{total heat flux}) into (\ref{total energy budget}) and taking into account the flow symmetries, we arrive at
\begin{eqnarray}
\label{total energy budget symmetry}
\displaystyle
PrRa(Nu-1)=\frac{\epsilon}{\kappa^3/H^4}
+\frac{2}{\beta (\kappa/H)^3}\left. {\left< w^{2} \right>}_{xyt} \right|_{z=0}
-\frac{1}{(\kappa/H)^3}\left. {\left< w^3 \right>}_{xyt} \right|_{z=0}. \nonumber \\
\end{eqnarray}
Note that in the impermeable case, i.e. conventional RBC, the energy budget is given by}
\begin{equation}
\label{budget RBC}
PrRa(Nu-1)=\frac{\epsilon}{\kappa^3/H^4}.
\end{equation}

\section{Scaling properties and turbulence structure}\label{sec:result}

%\subsection{$Nu$-$Ra$ scaling}\label{sec:nu-ra}

\GK{Let us first discuss the scaling property of the Nusselt number $Nu$ with the Rayleigh number $Ra$.
Figure \ref{fig:nu_ra} shows $Nu$ as a function of $Ra$.
It can be seen that the wall permeability leads to significant} heat transfer enhancement over the entire $Ra$ range.
In the impermeable case \GK{$\beta U=0$} the present DNS data in the horizontally-periodic domain \GK{are good agreement with the turbulent data obtained from the experiments \citep{chavanne01,niemela06} and the numerical simulation \citep{stevens10} performed} in cylindrical containers.
\GK{At high Rayleigh number $Ra\sim10^{8}$--$10^{10}$,} $Nu$ can be seen to scale with $Ra$ as $Nu\sim Ra^{1/3}$, nearly consistent with the well-known \GK{turbulence} scaling \citep[see e.g.][]{he12}.
In the permeable case \GK{$\beta U=3$}, on the other hand, the ultimate scaling $Nu\sim Ra^{1/2}$ can be observed at higher Rayleigh number \GK{$Ra\sim10^{7}$--$10^{10}$}, whereas the ordinary scaling $Nu\sim Ra^{1/3}$ is confirmed at lower \GK{Rayleigh number} $Ra\sim 10^{6}$--$10^7$.
It is worthy \GK{to} note that the scaling property of $Nu$ critically changes \GK{around} $Ra\sim 10^{7}$ from $Nu\sim Ra^{1/3}$ to $Nu\sim Ra^{1/2}$ with increasing $Ra$.

\begin{figure}
%\centering
\hspace*{.05\linewidth}
\begin{minipage}{.49\linewidth}
(\textit{a})\\
\includegraphics[clip,width=.8\linewidth]{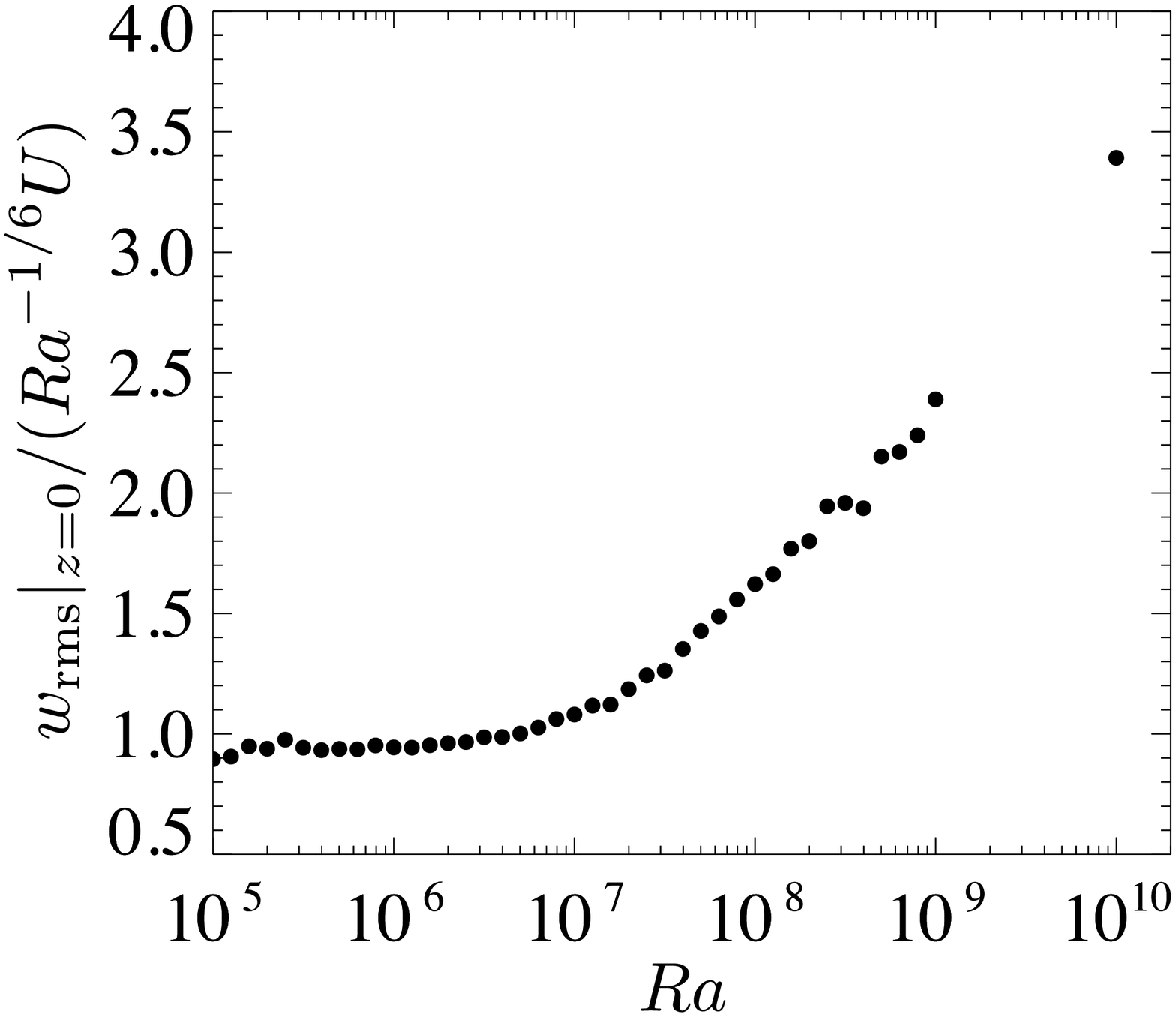}
\end{minipage}	
\begin{minipage}{.49\linewidth}
(\textit{b})\\
\includegraphics[clip,width=.8\linewidth]{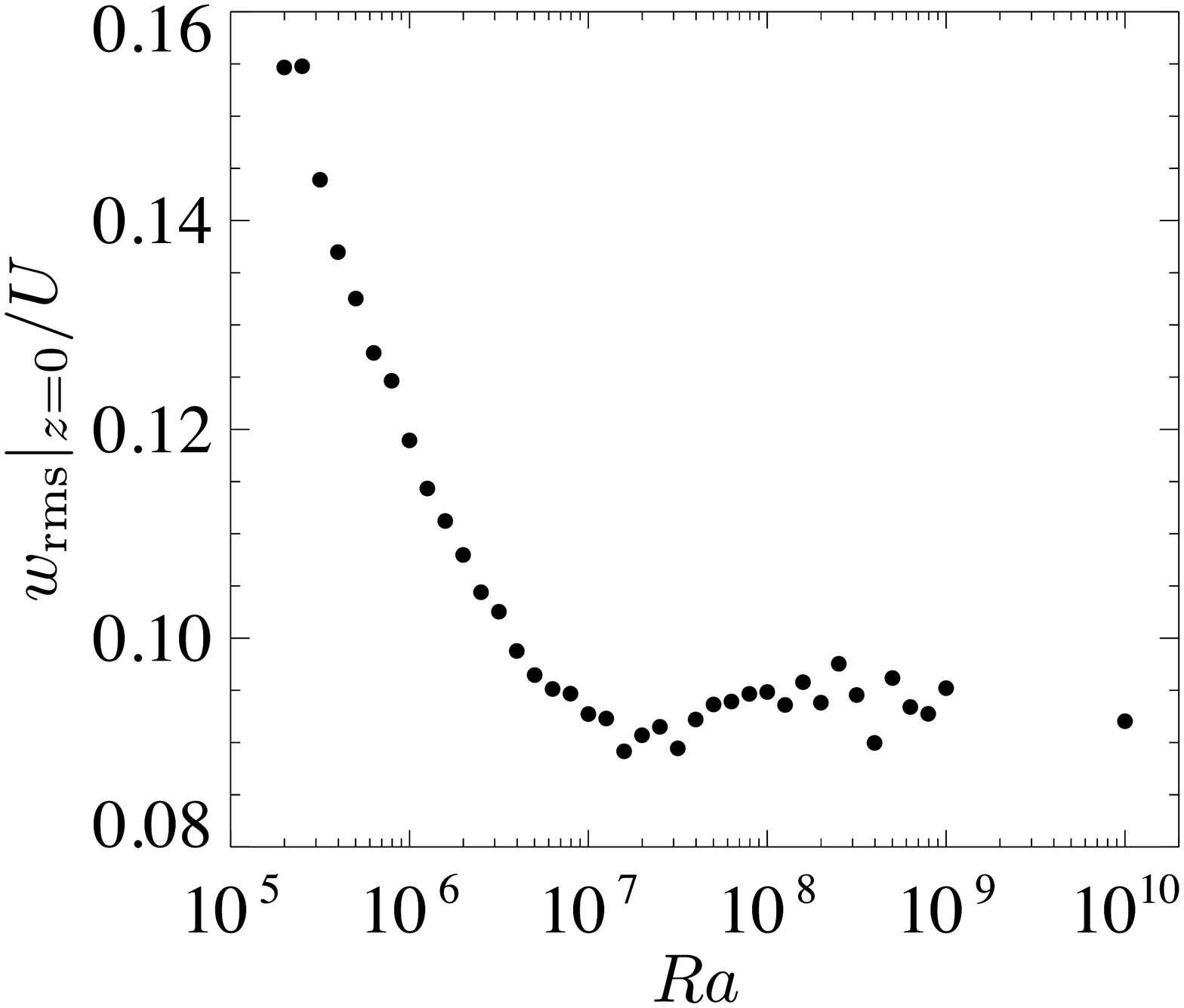}
\end{minipage}

\caption{The root-mean-square (RMS) vertical velocity on the wall $z=0$ normalised by (\textit{a})~\GK{$Ra^{-1/6}U$} and (\textit{b}) $U$ in the permeable case $\beta U=3$.
\label{fig:ra_wwall}}
\end{figure}

\begin{figure}
\centering
\begin{minipage}{.49\linewidth}
(\textit{a})\\
\includegraphics[clip,width=\linewidth]{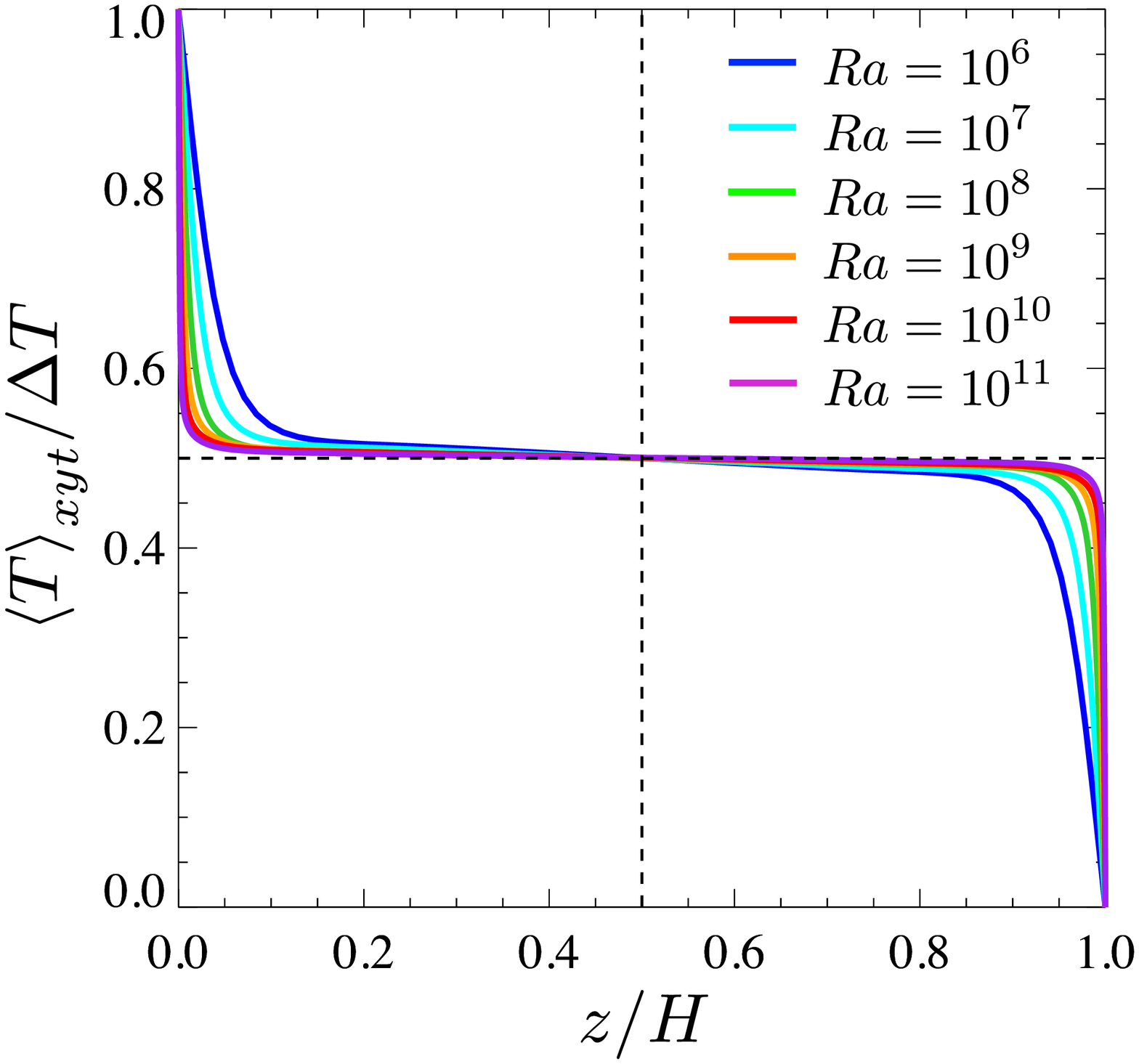}
\end{minipage}	
\begin{minipage}{.49\linewidth}
(\textit{b})\\
\includegraphics[clip,width=\linewidth]{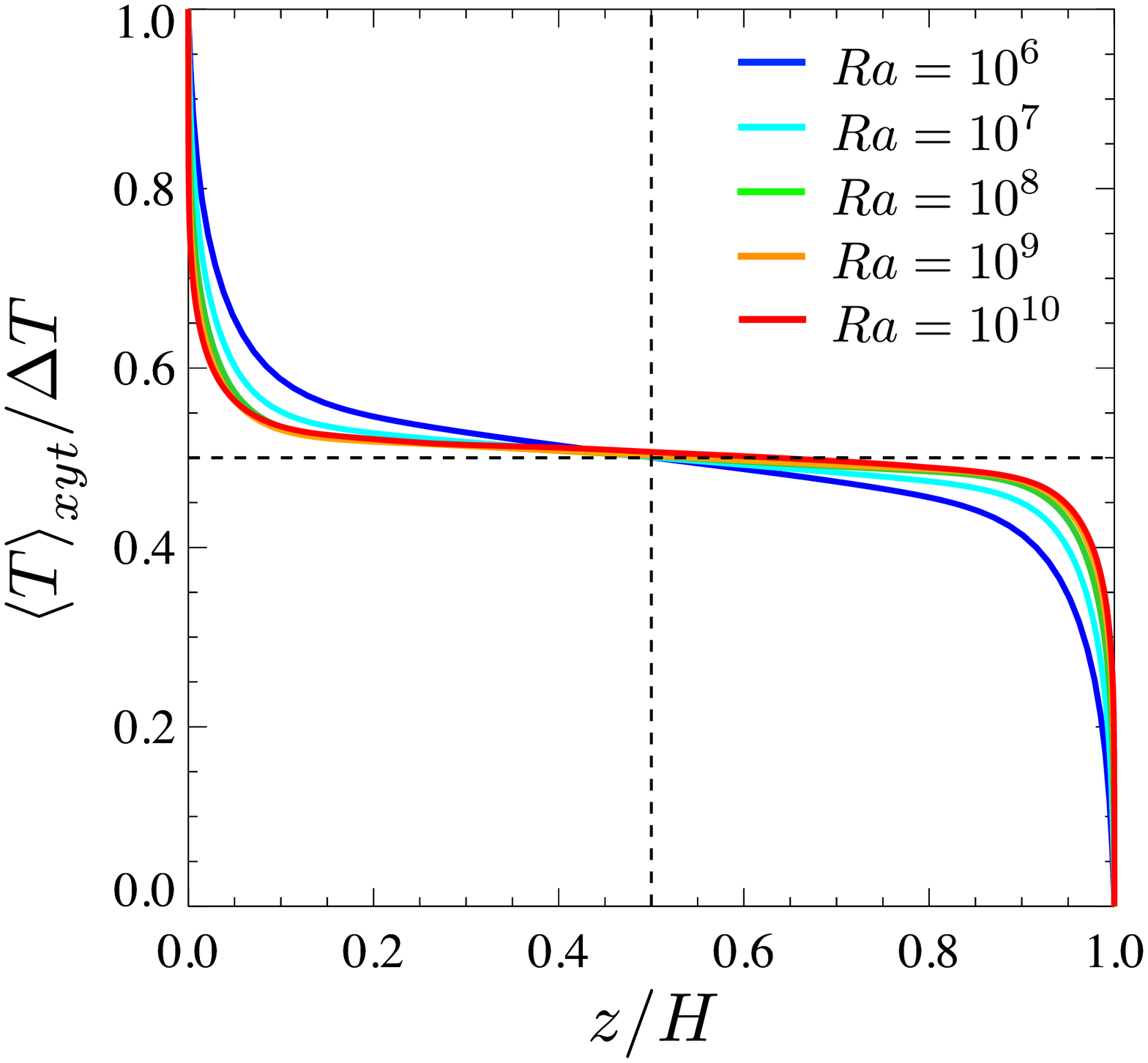}
\end{minipage}

\begin{minipage}{.49\linewidth}
(\textit{c})\\
\includegraphics[clip,width=\linewidth]{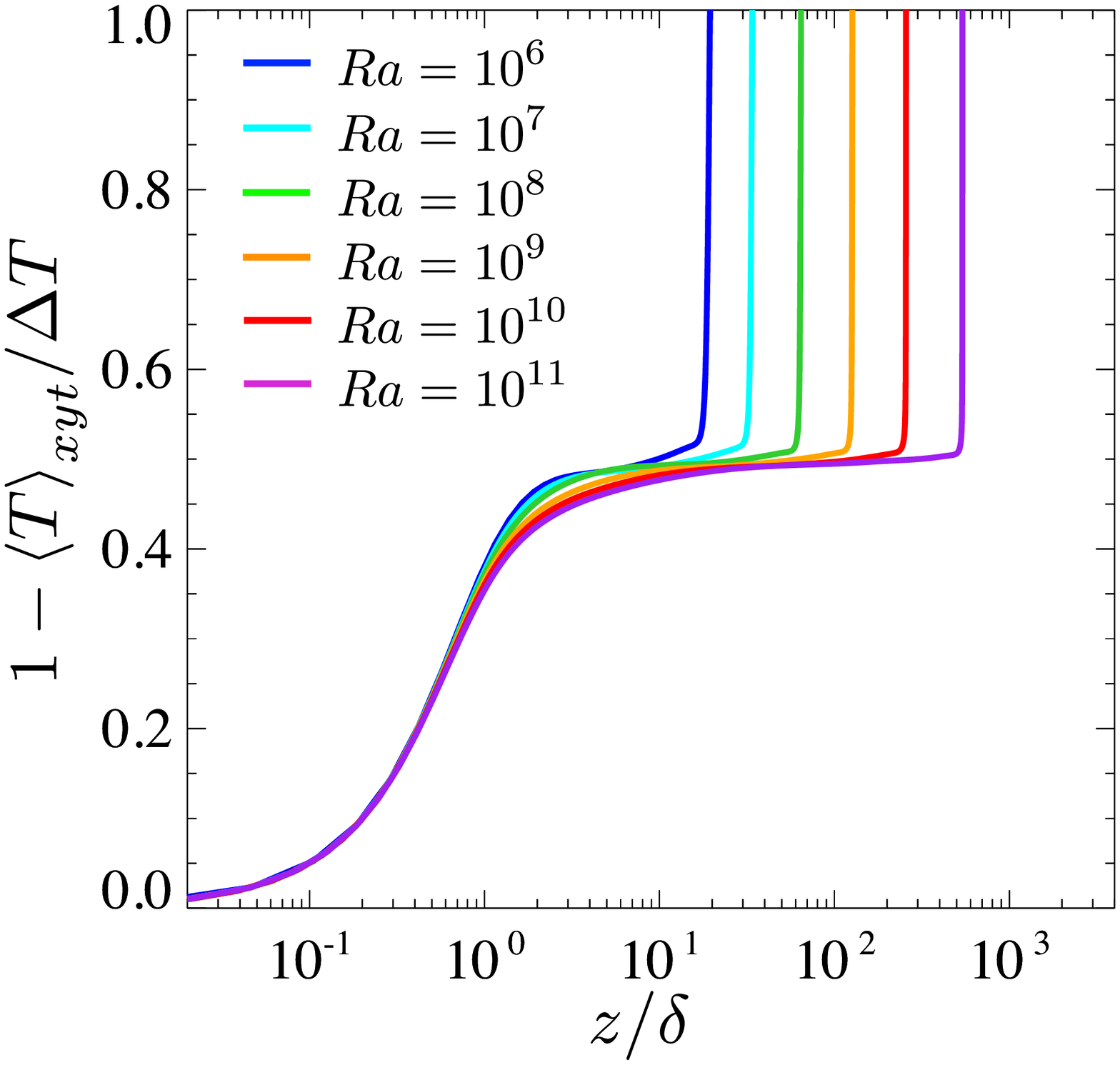}
\end{minipage}	
\begin{minipage}{.49\linewidth}
(\textit{d})\\
\includegraphics[clip,width=\linewidth]{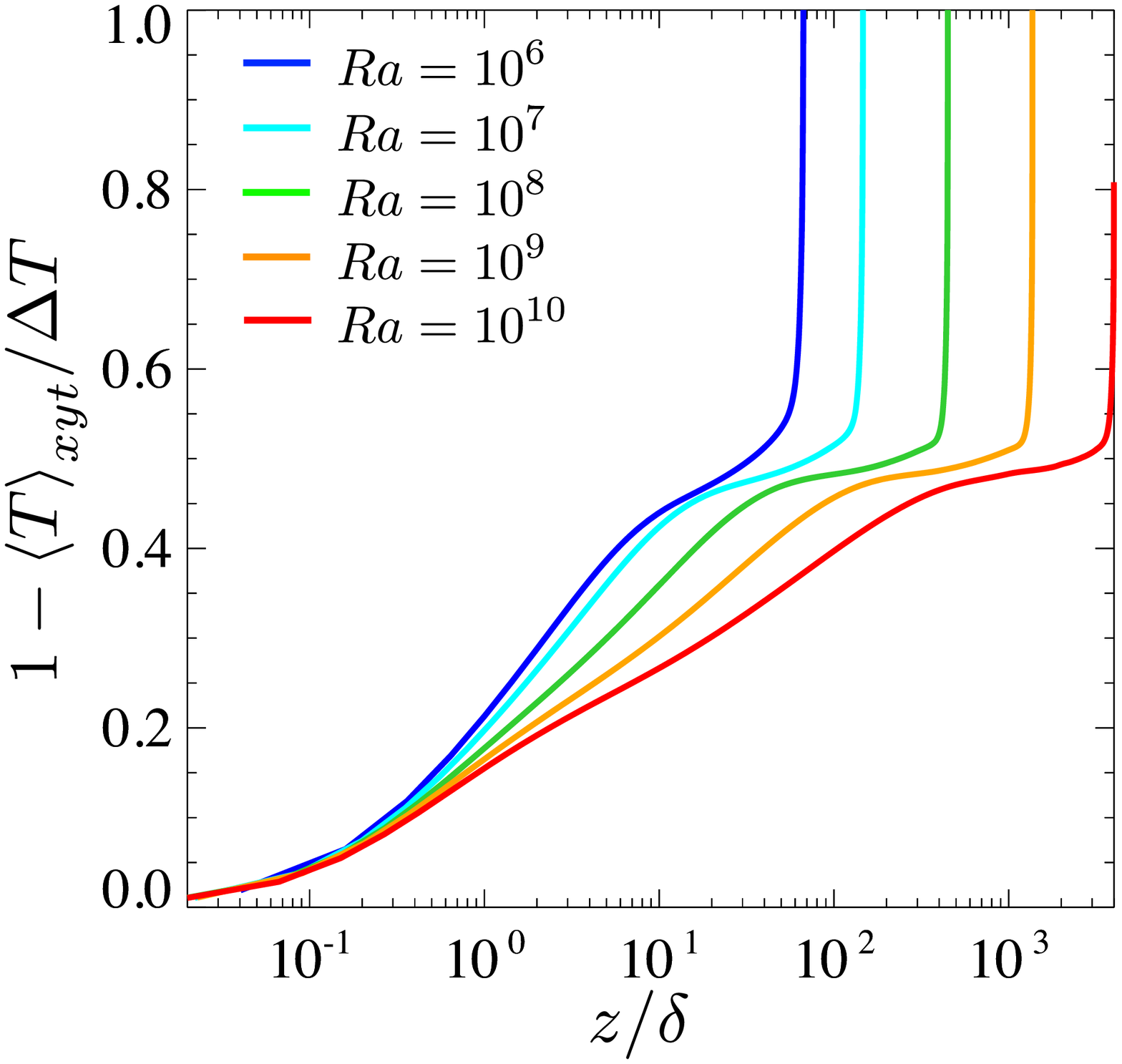}
\end{minipage}
\caption{Mean temperature profiles as a function of (\textit{a,b}) $z/H$ and (\textit{c,d}) $z/\delta$.
(\textit{a,c}) the impermeable case $\beta U=0$.
(\textit{b,d}) the permeable case $\beta U=3$.
\label{fig:te}}
\end{figure}

\begin{figure}
\centering
\begin{minipage}{.49\linewidth}
(\textit{a})\\
\includegraphics[clip,width=\linewidth]{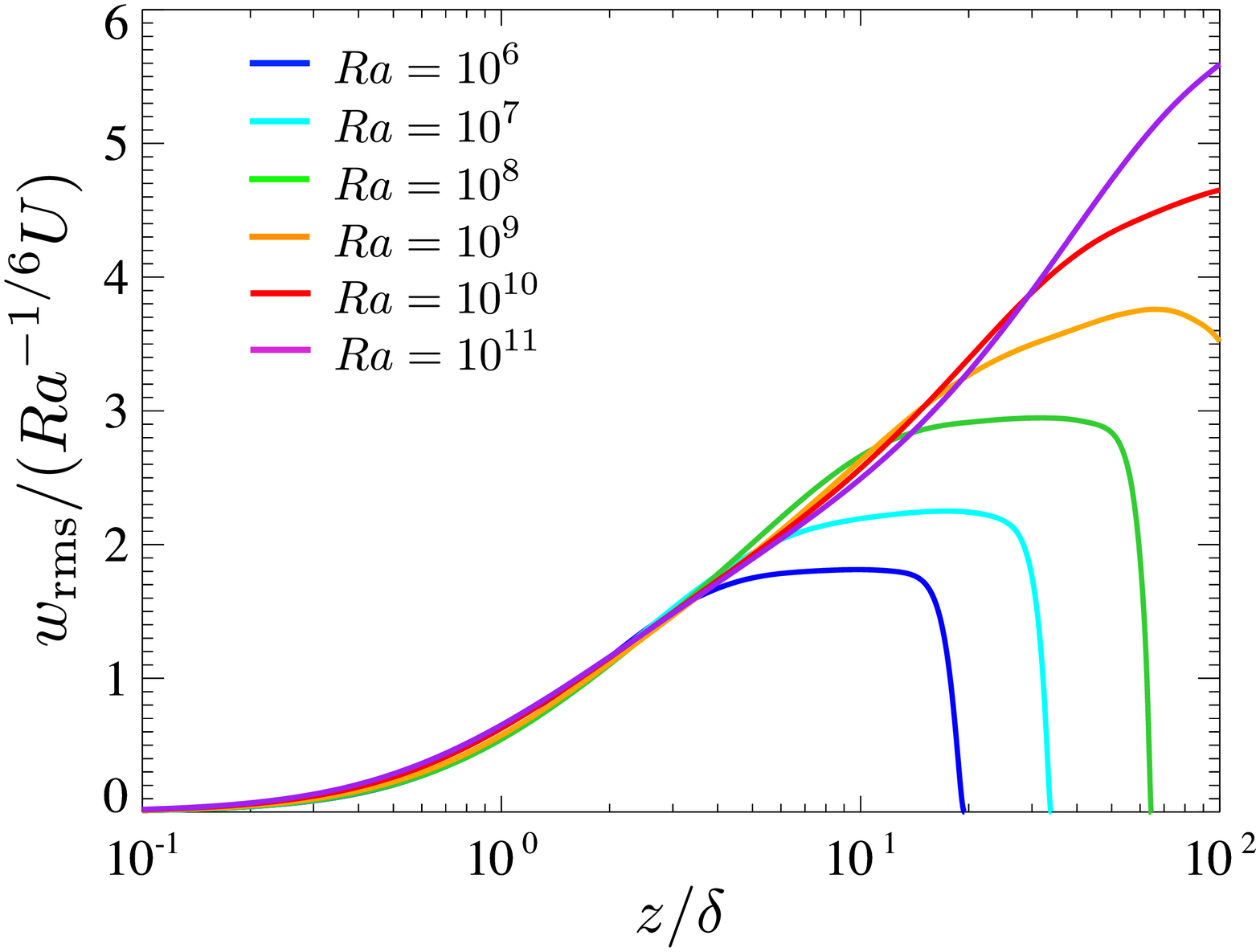}
\end{minipage}	
\begin{minipage}{.49\linewidth}
(\textit{b})\\
\includegraphics[clip,width=\linewidth]{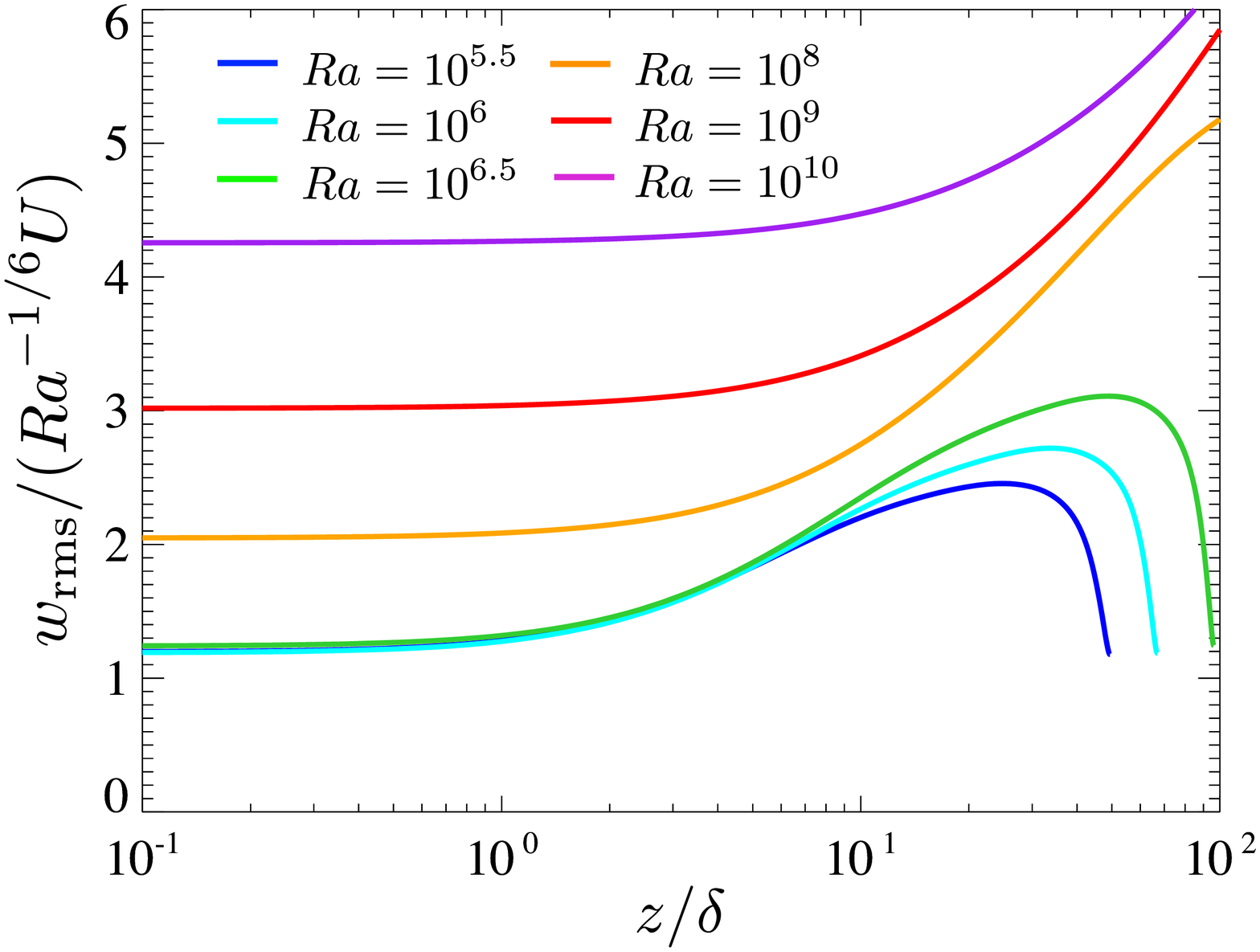}
\end{minipage}
\caption{The RMS vertical velocity normalised by \GK{$Ra^{-1/6} U$ as a function of $z/\delta$}.
({\it a}) the impermeable case $\beta U=0$.
({\it b}) the permeable case $\beta U=3$.
\label{fig:wrms_rbc_wall}}
\end{figure}

\GK{This} critical transition in the permeable case \GK{can also be confirmed undoubtedly} for the root-mean-square (RMS) vertical velocity \GK{$w_{{\rm rms}}=\langle w^{2} \rangle_{xyt}^{1/2}$} on the wall as shown in figure \ref{fig:ra_wwall}.
In the subcritical $Ra$ range $10^{6}\lesssim Ra\lesssim10^{7}$, \GK{the wall-normal transpiration velocity is weak in the sense that it is of the order of  $Ra^{-1/6} U$ (see figure \ref{fig:ra_wwall}\textit{a}), corresponding to the vertical velocity scale in the near-wall region of RBC for $Pr\sim 1$, i.e. the impermeable case, in  which the ordinary scaling $Nu \sim Ra^{1/3}$ has been observed.}
In the supercritical $Ra$ range $10^{7}\lesssim Ra\lesssim10^{10}$, on the other hand, the RMS velocity \GK{on the wall is significantly strong in the sense that it scales with the buoyancy-induced terminal velocity $U$ (see figure \ref{fig:ra_wwall}\textit{b}).
In \S\ref{sec:mechanism}, for the case of $Pr\sim 1$, the near-wall vertical velocity scale $Ra^{-1/6} U$ will be related with the ordinary scaling $Nu \sim Ra^{1/3}$, and the relevance of the vertical velocity scale $U$ to the ultimate scaling $Nu\sim Ra^{1/2}$ will also be discussed.}

\GK{We would like to stress that the ultimate heat transfer is not simplistically a consequence of just the wall permeability.
As will be shown later in this section, the wall permeability can trigger a critical change in convection states, consequently leading to the ultimate scaling $Nu\sim Ra^{1/2}$.}

%\subsection{Mean temperature}\label{sec:temperature}

\GK{Next we differentiate mean temperature profiles between the supercritical permeable case $\beta U=3$ at $Ra\gtrsim 10^7$ and the impermeable case $\beta U=0$.}
Figure \ref{fig:te} presents the mean temperature profiles \GK{in the impermeable and permeable cases}.
\GK{In the impermeable case, at higher $Ra$ the profile becomes flatter in the bulk region, while the near-wall temperature gradient becomes steeper.
In short the mean temperature profile $\langle T \rangle_{xyt}/\Delta T$ cannot scale with $z/H$.
The behaviour of the mean temperature in the subcritical case at $Ra\lesssim 10^7$ similar to that in the impermeable case.
In contrast to the impermeable case and the subcritical case, the mean temperature profile in the bulk region seems to scale with $\Delta T$ as a function of $z/H$ in the supercritical permeable case at $Ra\gtrsim 10^7$, and there remains a finite value of the temperature gradient, i.e., the order of $\Delta T/H$, therein even at high $Ra$.
This contrast should be a crucial consequence of the ultimate heat transfer as will be discussed in \S\ref{sec:mechanism}.} 

\begin{figure}
	\centering
	\begin{minipage}{.32\linewidth}
		(\textit{a})\\
		\includegraphics[clip,width=\linewidth]{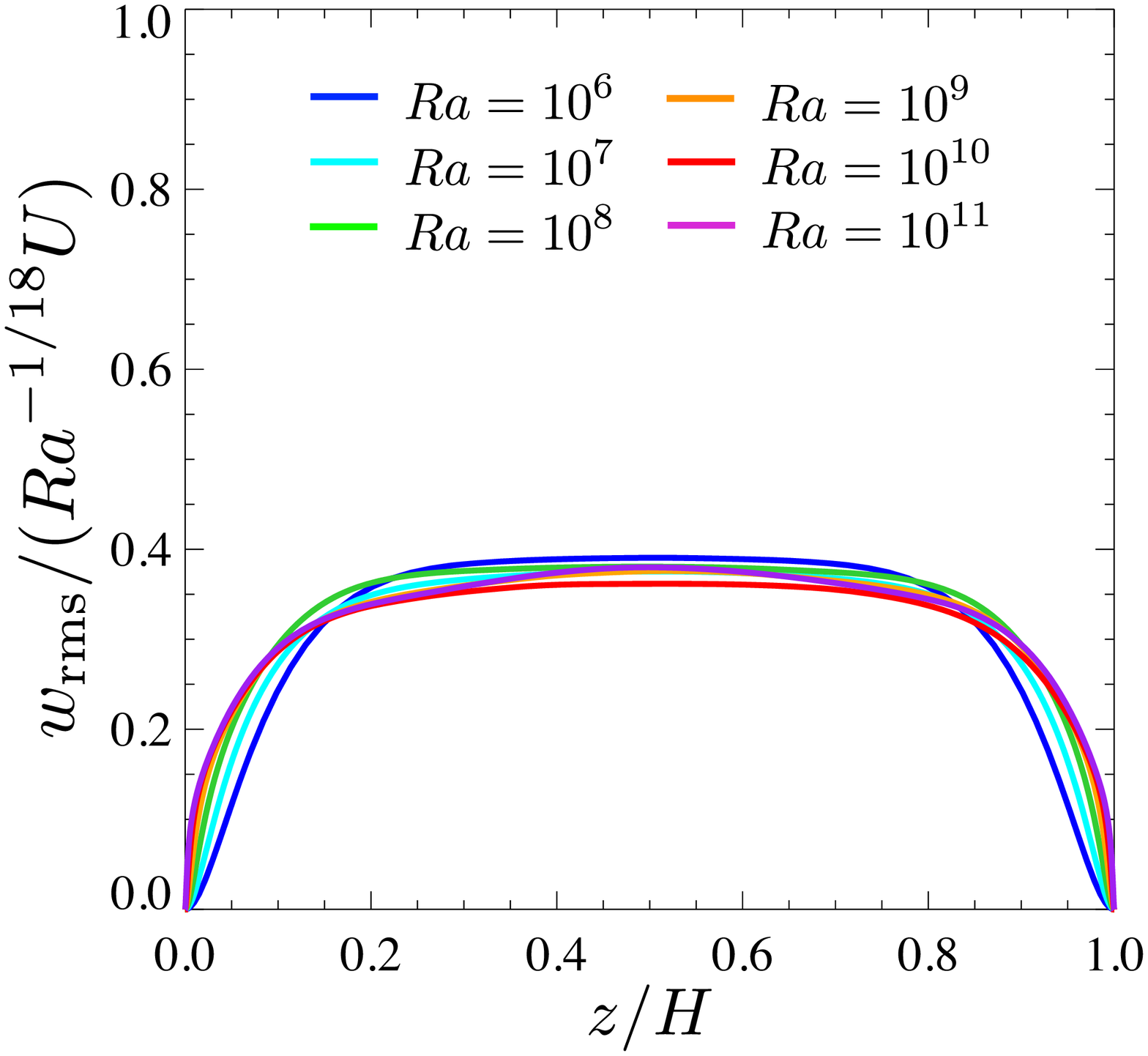}
	\end{minipage}
	\begin{minipage}{.32\linewidth}
		(\textit{b})\\
		\includegraphics[clip,width=\linewidth]{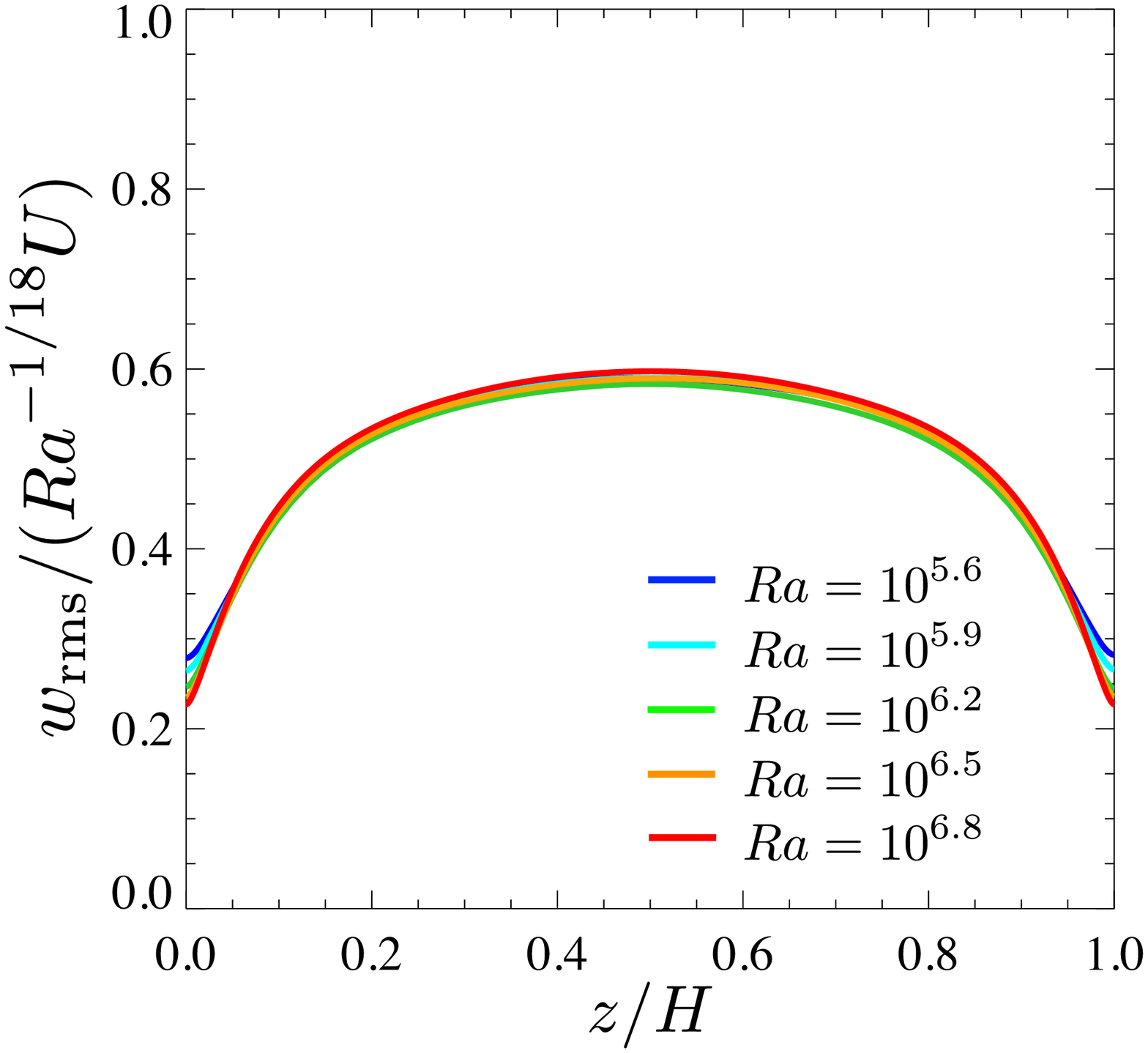}
	\end{minipage}
	\begin{minipage}{.32\linewidth}
		(\textit{c})\\
		\includegraphics[clip,width=\linewidth]{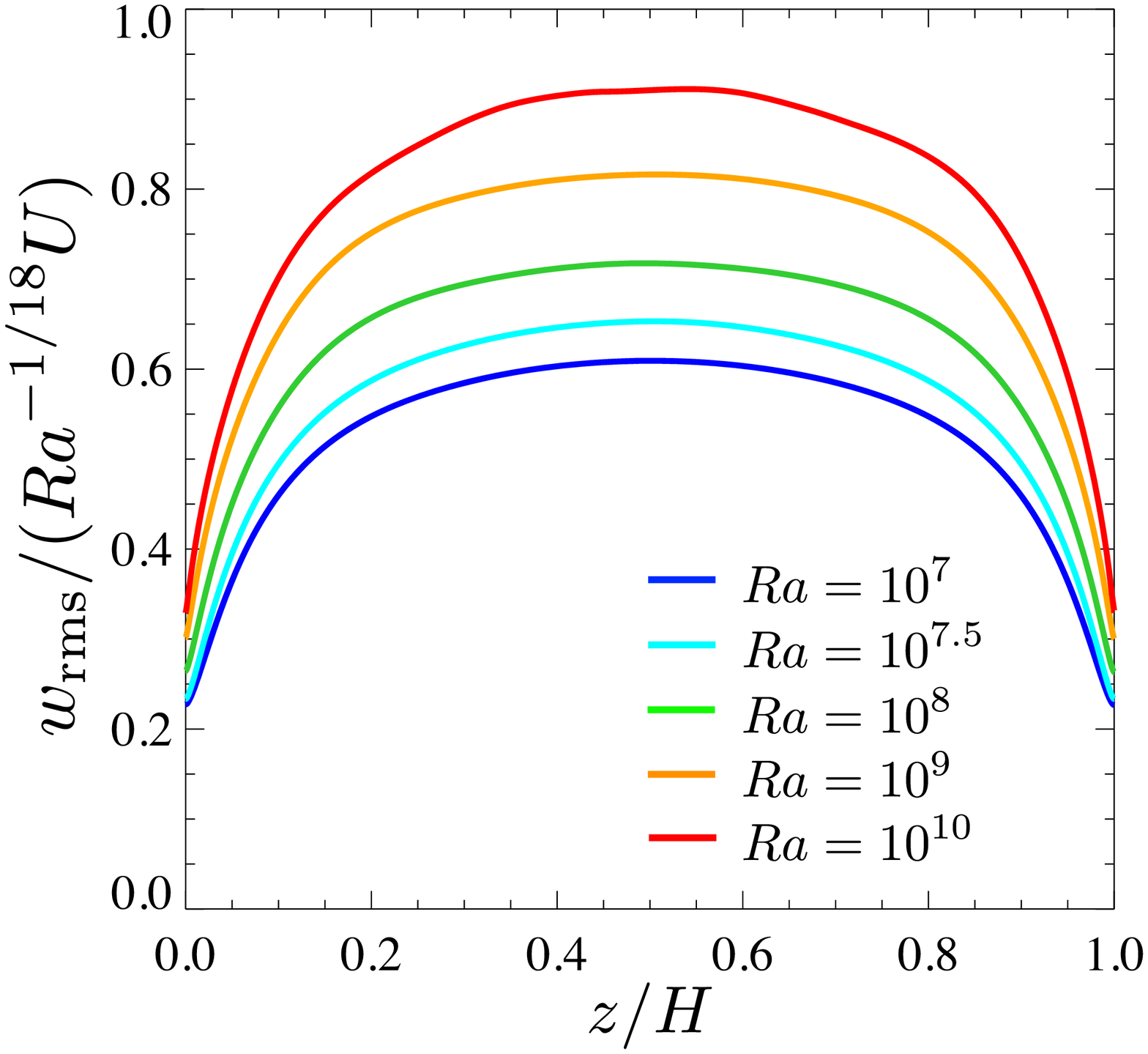}
	\end{minipage}
	
	\begin{minipage}{.32\linewidth}
		(\textit{d})\\
		\includegraphics[clip,width=\linewidth]{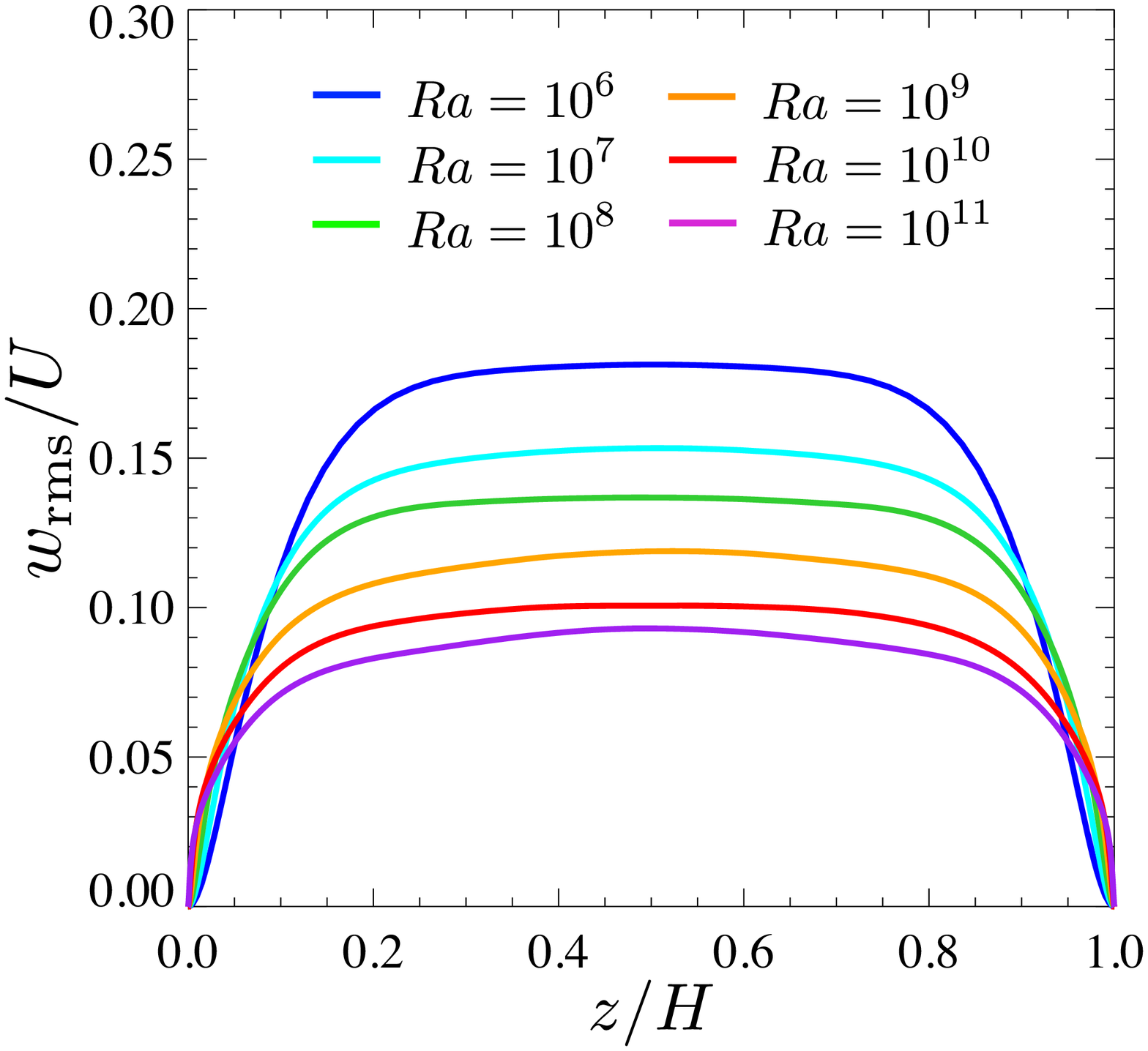}
	\end{minipage}
	\begin{minipage}{.32\linewidth}
		(\textit{e})\\
		\includegraphics[clip,width=\linewidth]{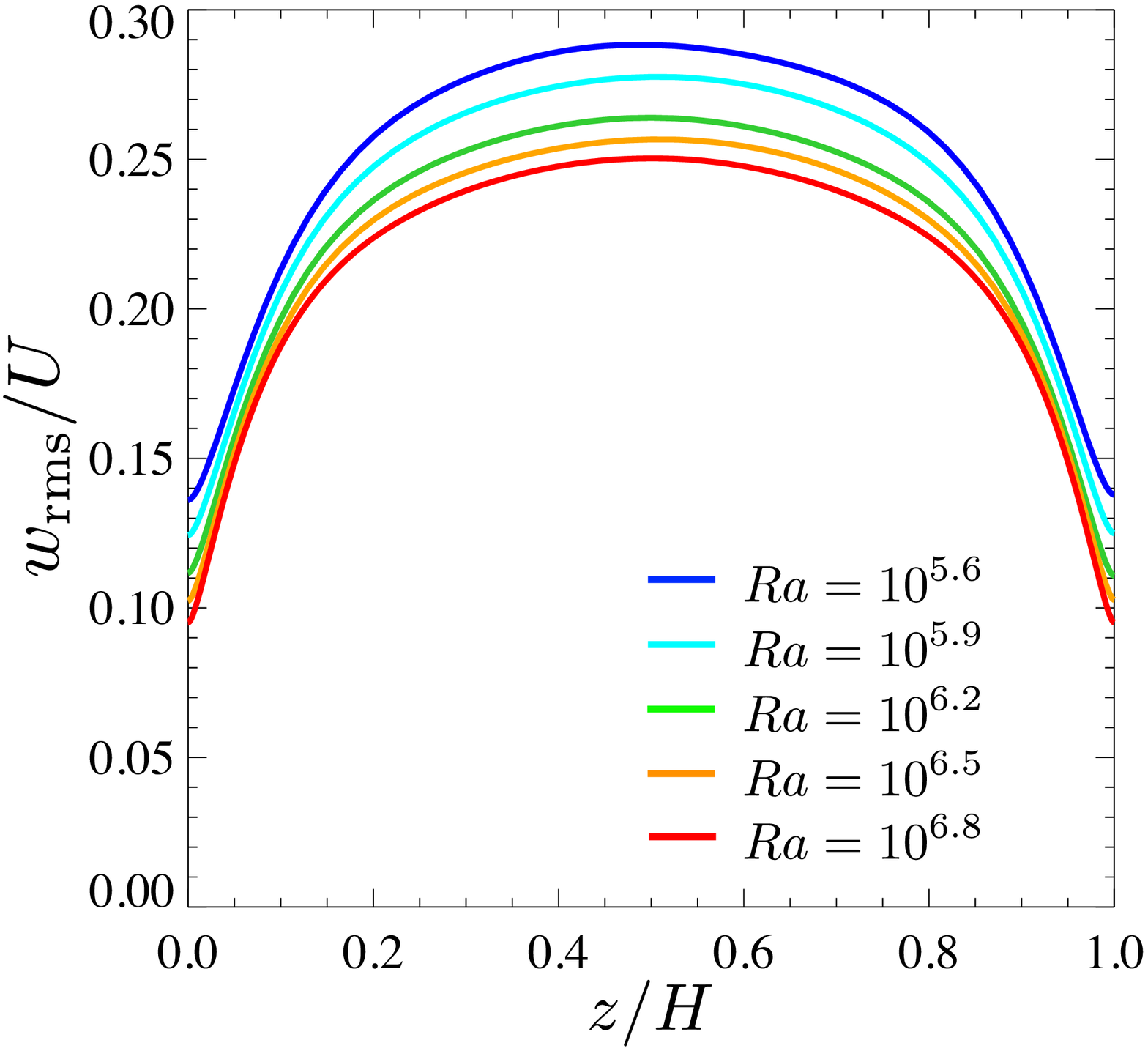}
	\end{minipage}
	\begin{minipage}{.32\linewidth}
		(\textit{f})\\
		\includegraphics[clip,width=\linewidth]{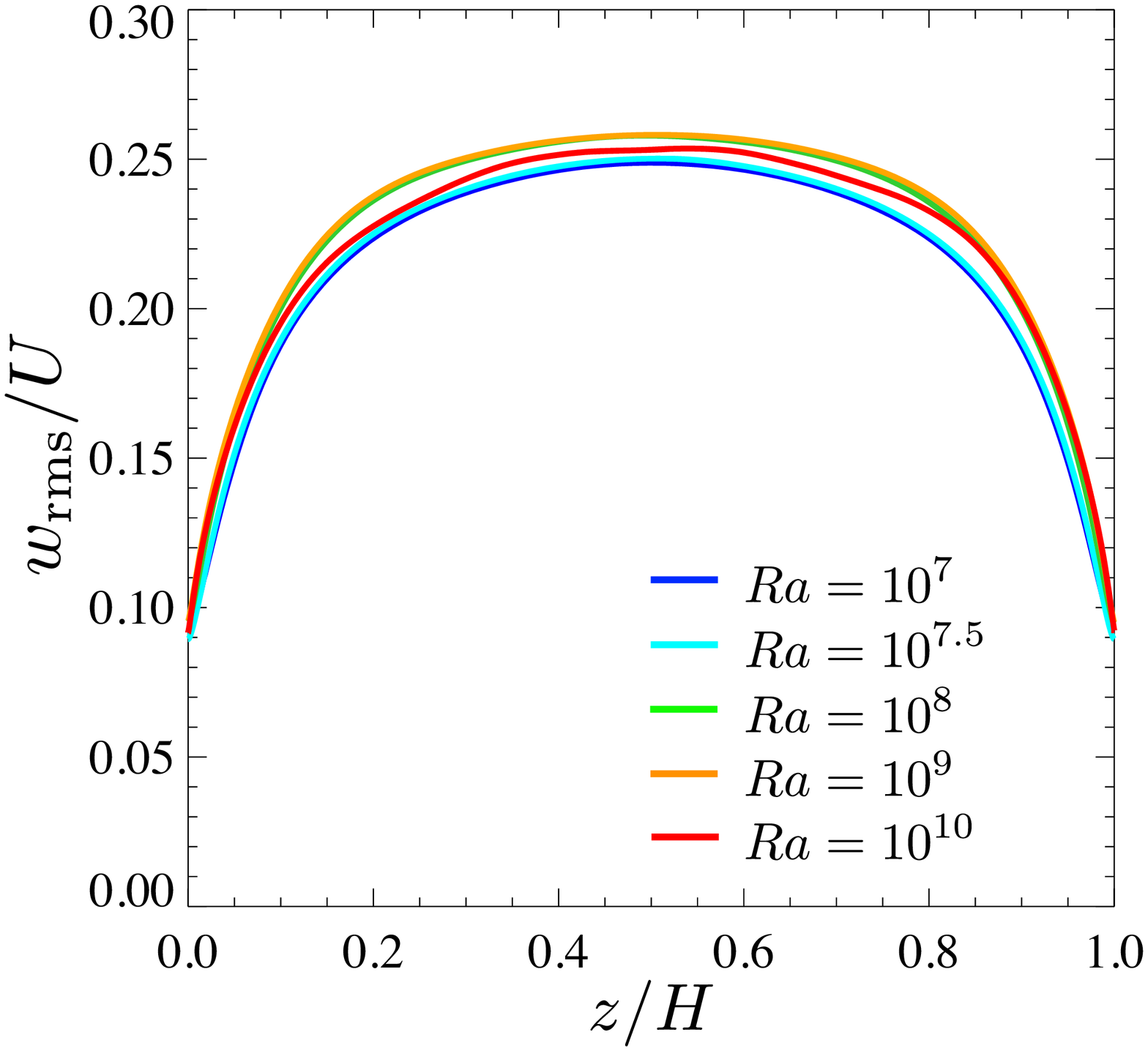}
	\end{minipage}
	\caption{The RMS vertical velocity normalised by \GK{(\textit{a--c}) $Ra^{-1/18} U$ and (\textit{d--f}) $U$.
			(\textit{a,d}) the impermeable case $\beta U=0$.
			(\textit{b,e}) the subcritical permeable case $\beta U=3$ at $10^{5.6}\le Ra\le10^{6.8}$.
			(\textit{c,f}) the supercritical permeable case $\beta U=3$ at $10^{7}\le Ra\le10^{10}$.}
		\label{fig:urms}}
\end{figure}

\begin{figure}
	\centering
	\begin{minipage}{.32\linewidth}
		(\textit{a})\\
		\includegraphics[clip,width=\linewidth]{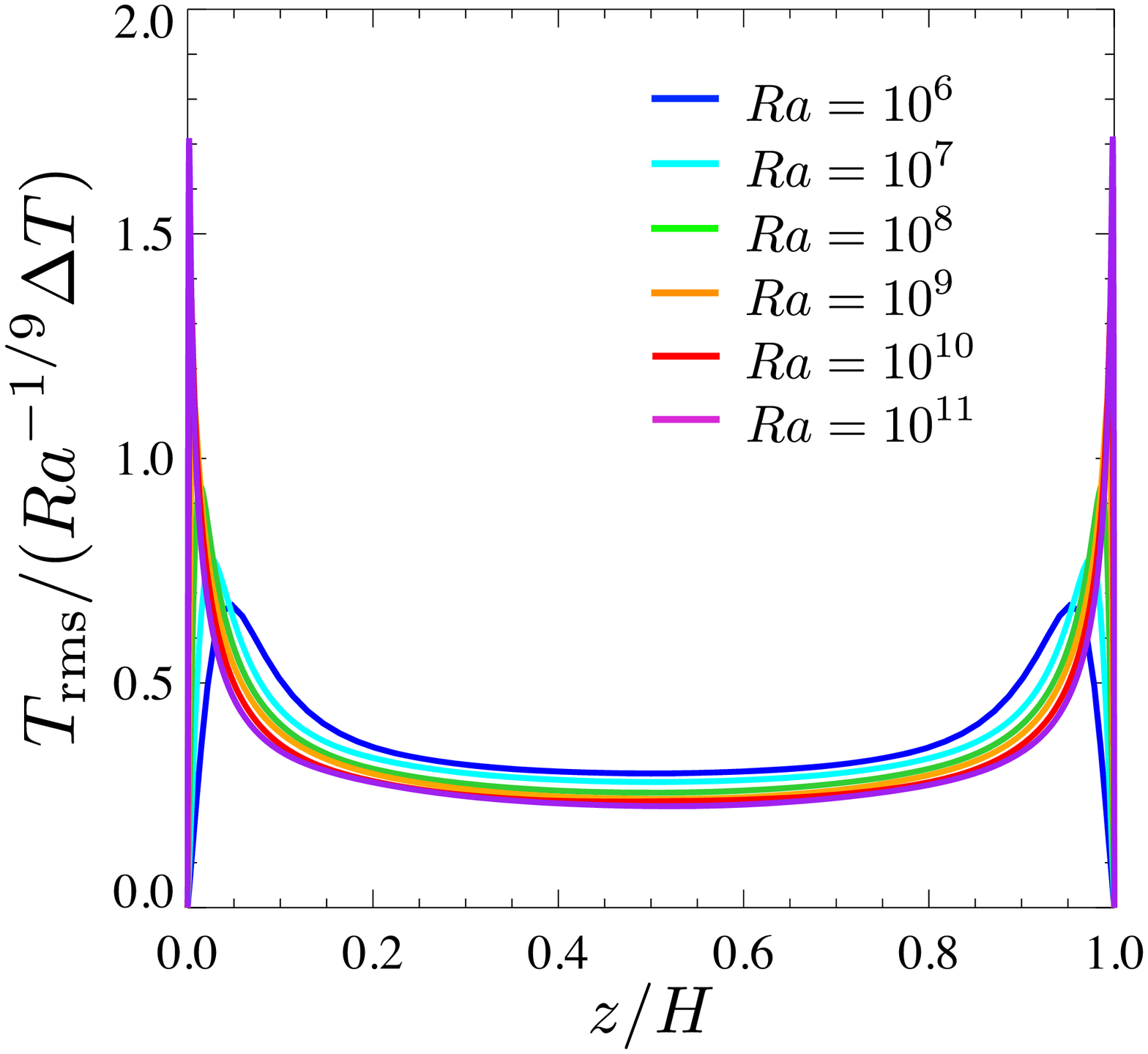}
	\end{minipage}
	\begin{minipage}{.32\linewidth}
		(\textit{b})\\
		\includegraphics[clip,width=\linewidth]{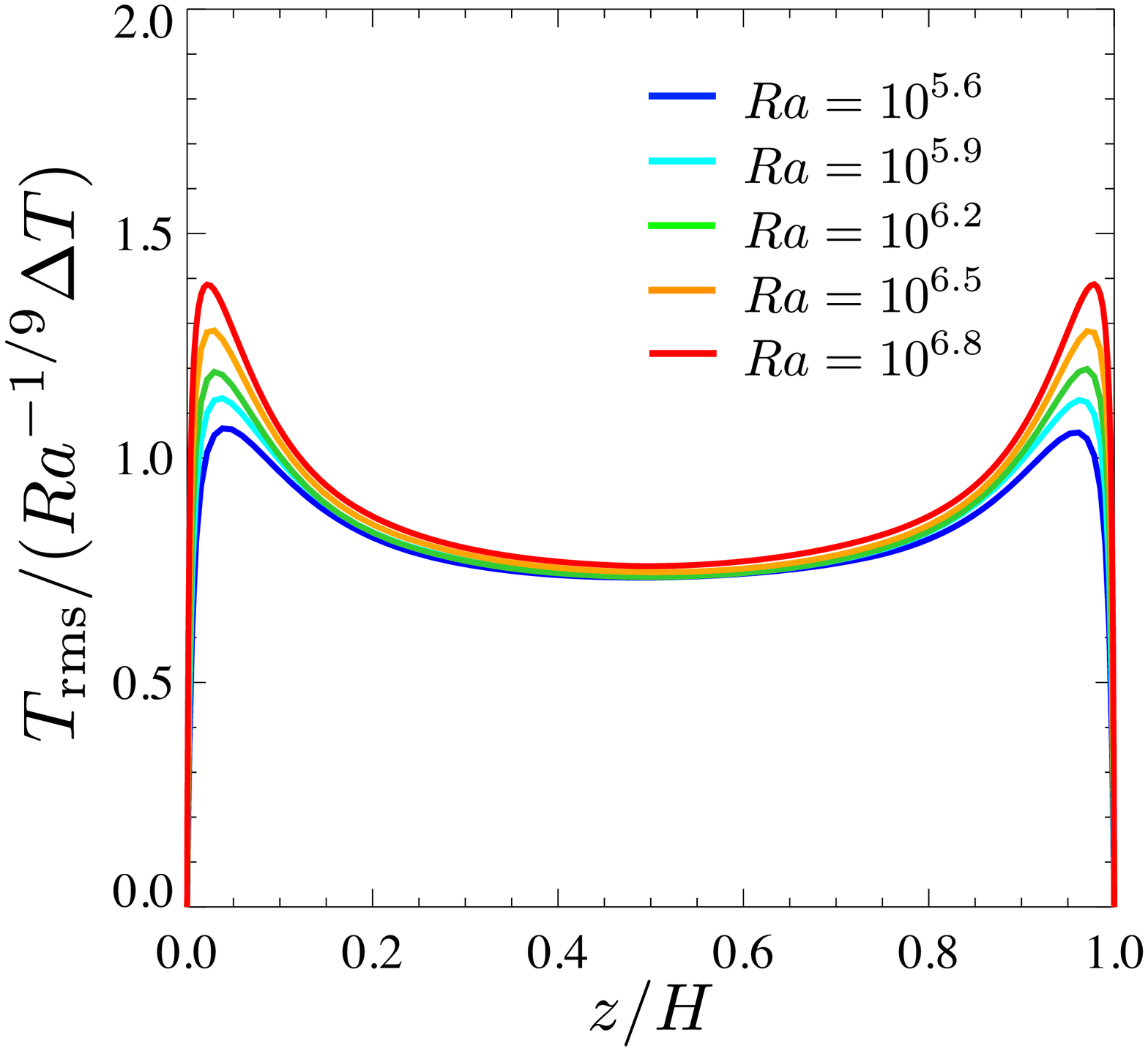}
	\end{minipage}
	\begin{minipage}{.32\linewidth}
		(\textit{c})\\
		\includegraphics[clip,width=\linewidth]{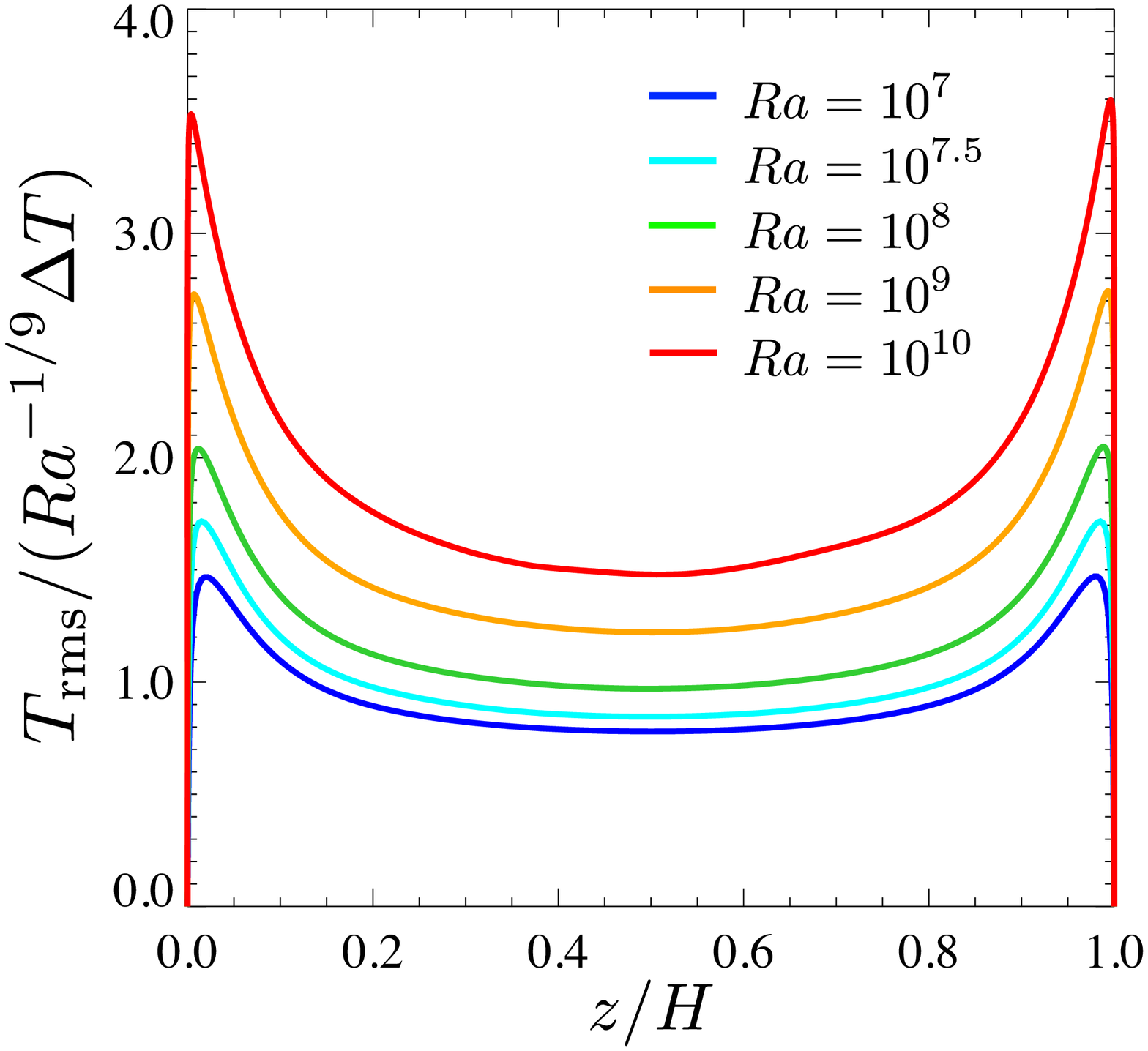}
	\end{minipage}
	
	\begin{minipage}{.32\linewidth}
		(\textit{d})\\
		\includegraphics[clip,width=\linewidth]{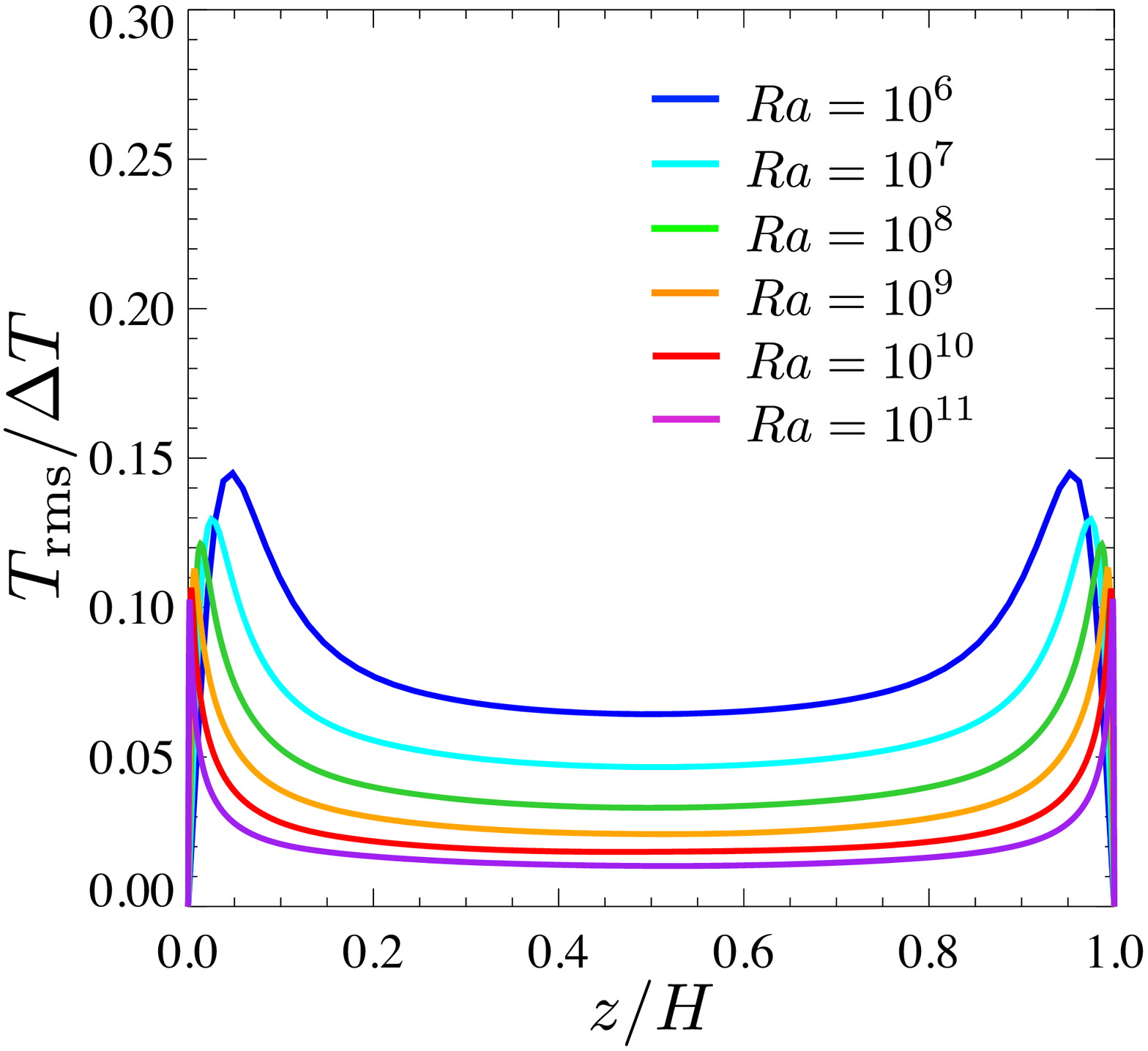}
	\end{minipage}
	\begin{minipage}{.32\linewidth}
		(\textit{e})\\
		\includegraphics[clip,width=\linewidth]{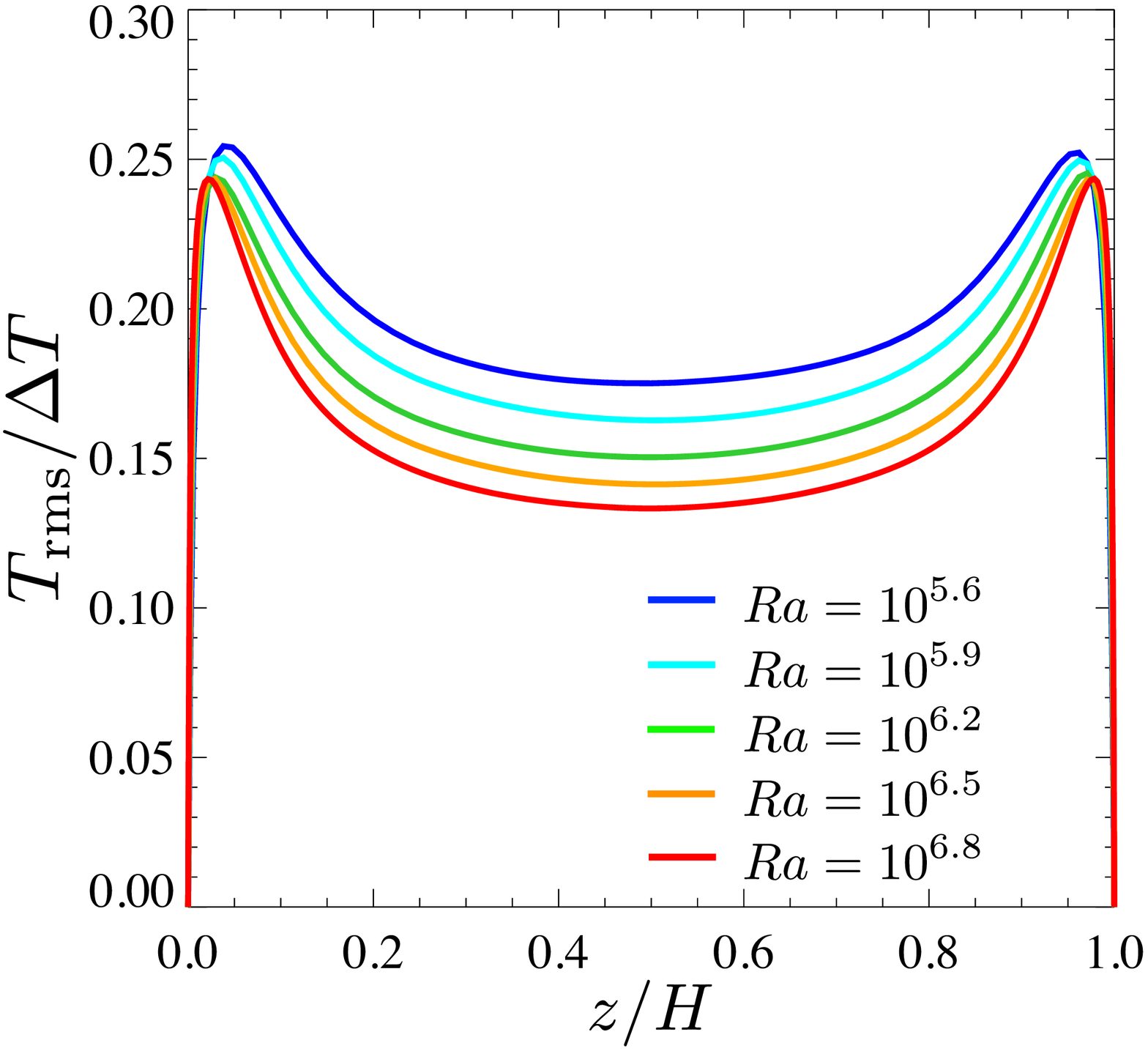}
	\end{minipage}
	\begin{minipage}{.32\linewidth}
		(\textit{f})\\
		\includegraphics[clip,width=\linewidth]{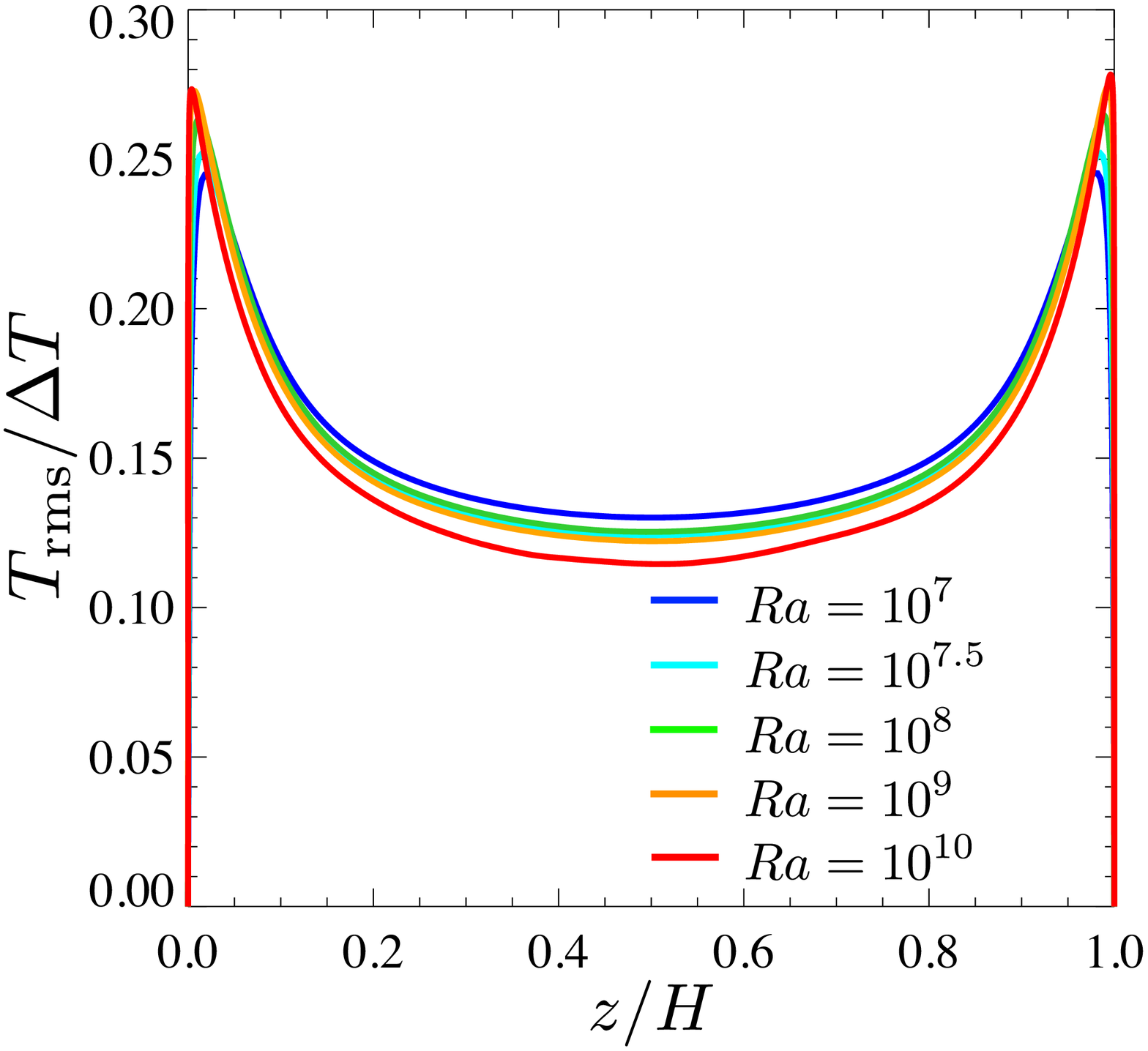}
	\end{minipage}
	\caption{The RMS \GK{temperature normalised by (\textit{a--c}) $Ra^{-1/9} \Delta T$ and (\textit{d--f}) $\Delta T$.
			(\textit{a,d}) the impermeable case $\beta U=0$.
			(\textit{b,e}) the subcritical permeable case $\beta U=3$ at $10^{5.6}\le Ra\le10^{6.8}$.
			(\textit{c,f}) the supercritical permeable case $\beta U=3$ at $10^{7}\le Ra\le10^{10}$.}
		\label{fig:terms}}
\end{figure}

In the permeable case with isothermal wall boundaries, different from the thermal convection without horizontal walls \citep{calzavarini05,pawar16}, \GK{there exists a thermal conduction layer on the wall, where heat transfer by conduction dominates over that by convection.
In figure \ref{fig:te}(\textit{c,d}) are shown the mean temperature profiles $1-\langle T \rangle_{xyt}/\Delta T$ as a function of $z/\delta$, where $\delta$ is the thickness of a thermal conduction layer defined as}
\begin{equation}
\GK{\delta\equiv}-\Delta T{\left( \left. \frac{{\rm d}{\left< T \right>}_{xyt}}{{\rm d}z} \right|_{z=0} \right)}^{-1}=\frac{H}{2 Nu}.
\label{eq_delta}
\end{equation}
All the profiles in the impermeable case collapse \GK{onto} a single curve in the thermal \GK{conduction} layer $z/\GK{\delta}\lesssim 1$.
\GK{It is also the case in the permeable case; however, the thermal conduction cannot be dominant around $z/\delta\sim 1$ where the convection is also important.
The large difference of the temperature profiles at $z/\delta\gtrsim 1$ in the supercritical permeable case at $Ra\gtrsim 10^7$ implies its reasonable scaling with $z/H$, shown in figure~\ref{fig:te}{\it b}.}

%\subsection{RMS velocity and temperature}\label{sec:rms}

\GK{As mentioned before, the vertical velocity fluctuation on the permeable walls scales with $Ra^{-1/6} U$ at subcritical Rayleigh number $Ra\lesssim 10^7$.
In the impermeable case (in addition to the subcritical permeable case) the near-wall RMS vertical velocity $w_{\rm rms}=\langle w^{2} \rangle_{xyt}^{1/2}$ also scales with $Ra^{-1/6} U$ as a function of $z/\delta$ (see figure \ref{fig:wrms_rbc_wall}{\it a}).
However, the vertical velocity fluctuation in the supercritical case at $Ra\gtrsim 10^7$ exhibits quite distinct behaviour from that in the impermeable and subcritical cases (see figure \ref{fig:wrms_rbc_wall}{\it b})}.

\GK{Figures \ref{fig:urms}({\it a--c},{\it d--f}) show the RMS vertical velocity normalised by the velocity scale $Ra^{-1/18} U$ and the buoyancy-induced terminal velocity $U$, respectively.
In the impermeable case at $10^{6}\le Ra\le 10^{11}$ and the subcritical permeable case at $10^{5.6}\le Ra\le10^{6.8}$, the RMS vertical velocity in the bulk region is seen to scale with $Ra^{-1/18} U$ corresponding to the vertical velocity scale in the bulk region of RBC for $Pr\sim 1$ (figure \ref{fig:urms}{\it a,b}), and thus it decreases relatively with respect to $U$ as $Ra$ increases (figure \ref{fig:urms}{\it d,e}).
In the supercritical permeable case at $10^{7}\le Ra\le10^{10}$, on the other hand, the velocity fluctuation in the bulk is found to scale with $U$ (figure \ref{fig:urms}{\it f}).
The near-wall gradient of $w_{\rm rms}$ with respect to $z/H$ in figure \ref{fig:urms} is steeper at higher $Ra$ in the impermeable and the subcritical permeable cases, but the same is not true of the supercritical permeable case.  
Although $w_{\rm rms}$ is not null on the permeable walls as already shown in figure~\ref{fig:ra_wwall}, the ratio of near-wall $w_{\rm rms}$ to bulk $w_{\rm rms}$ should be of the order of $Ra^{-1/9}$ in the subcritical case, implying that the near-wall vertical velocity fluctuation becomes smaller than that in the bulk region at higher $Ra$ (see figure~\ref{fig:urms}{\it b}).
The vertical RMS velocities $w_{\rm rms}/U$ as a function of $z/H$ are almost independent of the Rayleigh number $Ra$ in the whole region of the supercritical permeable case.}
Note that near the walls, the RMS velocity is suppressed due to the presence of the walls even in \GK{the supercritical permeable case exhibiting the ultimate scaling $Nu\sim Ra^{1/2}$.
Needless to say, such suppression of the vertical velocity has not been observed in the ultimate heat transfer in wall-less thermal convection
\citep{calzavarini05, pawar16}.}

The RMS temperature \GK{$T_{\rm rms}=\langle (T-{\langle T \rangle}_{xyt})^{2} \rangle_{xyt}^{1/2}$ normalised by the temperature scale $Ra^{-1/9} \Delta T$ and the temperature difference $\Delta T$ between the walls is shown in figures \ref{fig:terms}({\it a--c},{\it d--f}), respectively.
In the bulk region of the impermeable and subcritical permeable cases, the RMS temperature is seen to scale with $Ra^{-1/9}\Delta T$ (figure~\ref{fig:terms}{\it a,b}), and so it decreases as $Ra$ increases.
On the other hand, the temperature fluctuation in the supercritical permeable case is found to scale with $\Delta T$ (figure~\ref{fig:terms}{\it f}).
This remarkable difference in the scalings of the temperature fluctuation originates from the scaling difference in the mean temperature (cf. figure~\ref{fig:te}).
In the supercritical permeable case the vertical fluid motion across the sustaining mean temperature difference of $O(\Delta T)$ in the bulk region can induce the temperature fluctuation of $O(\Delta T)$ even at higher $Ra$, but in the impermeable and the subcritical cases the vanishing mean temperature difference means the small temperature fluctuation.
In \S\ref{sec:mechanism} we shall discuss the different scaling properties of the RMS vertical velocity with $Ra^{-1/18} U$ and $U$ as well as the difference in scaling of the temperature fluctuation with $Ra^{-1/9} \Delta T$ and $\Delta T$.}

\begin{figure}
	\centering
	\begin{minipage}{.49\linewidth}
		(\textit{a})\\
		\includegraphics[clip,width=\linewidth]{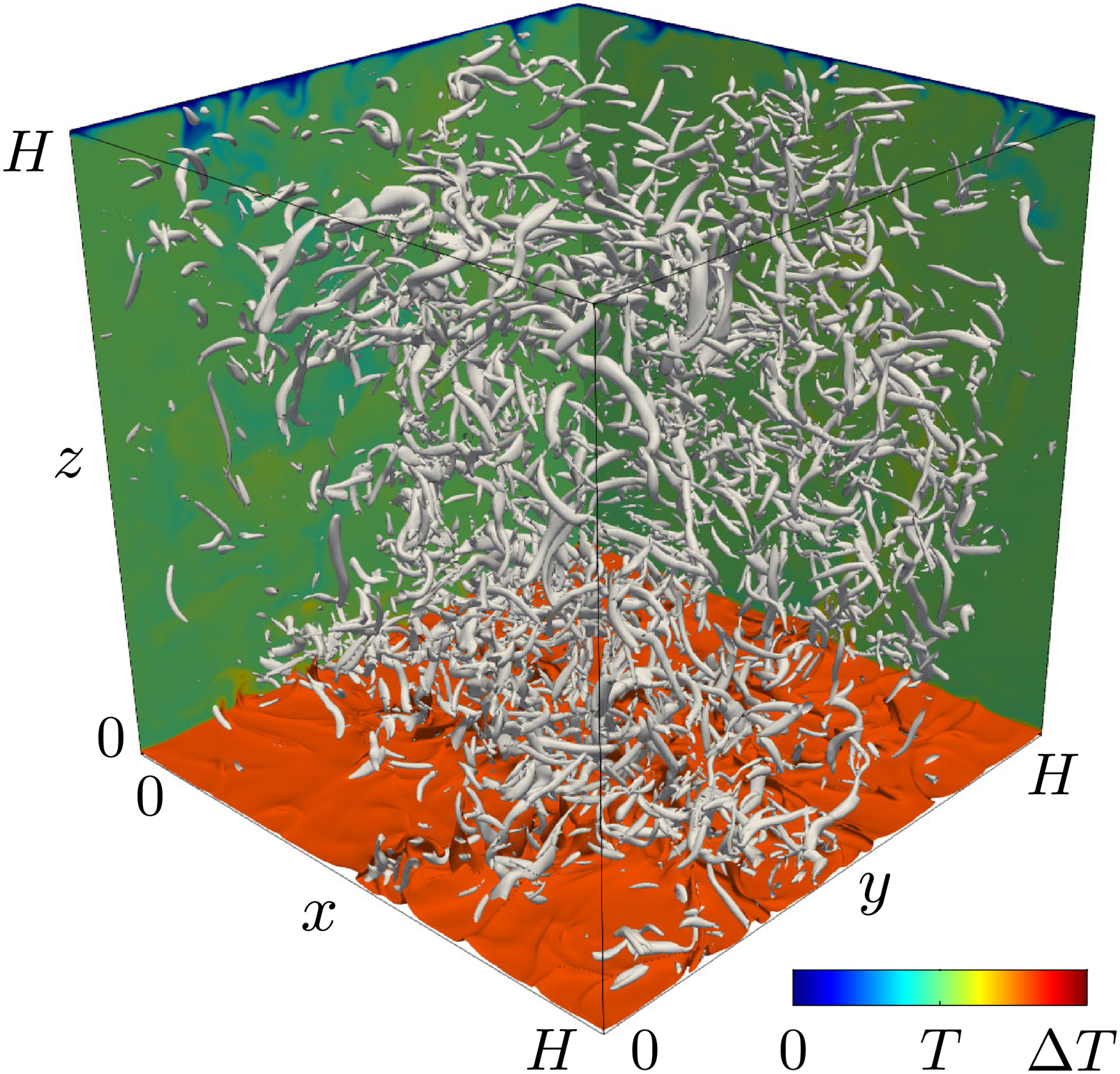}
	\end{minipage}
	\begin{minipage}{.49\linewidth}
		(\textit{b})\\
		\includegraphics[clip,width=\linewidth]{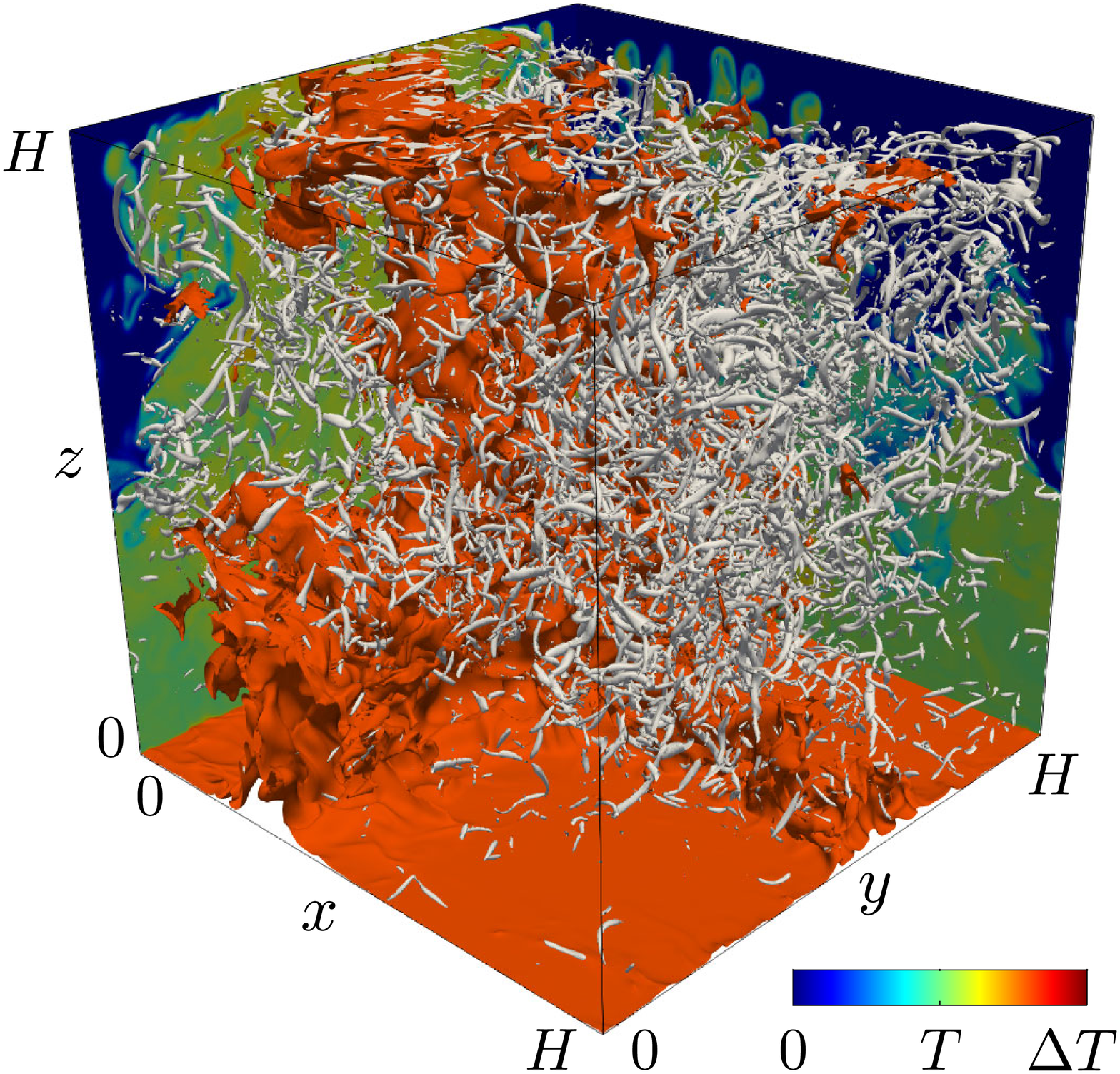}
	\end{minipage}
	\caption{Instantaneous thermal and \GK{vortical} structures in (\textit{a}) the impermeable case \GK{$\beta U=0$ and (\textit{b}) the supercritical permeable case $\beta U=3$} at $Ra=10^{9}$.
		The orange and grey objects respectively represent the isosurfaces of the temperature $T/\Delta T =0.7$ and \GK{of the second invariant of the velocity gradient tensor, (\textit{a}) $Q/(\nu^{2}/H^{4})=8\times10^{10}$ and (\textit{b}) $Q/(\nu^{2}/H^{4})=4.8\times10^{11}$.}
		\GK{The colour indicates the temperature distribution on the planes $x=0$ and $y=H$.}
		\label{fig:paraview}}
\end{figure}

\begin{figure}
	\centering
	\begin{minipage}{.49\linewidth}
		(\textit{a})\\
		\includegraphics[clip,width=.8\linewidth]{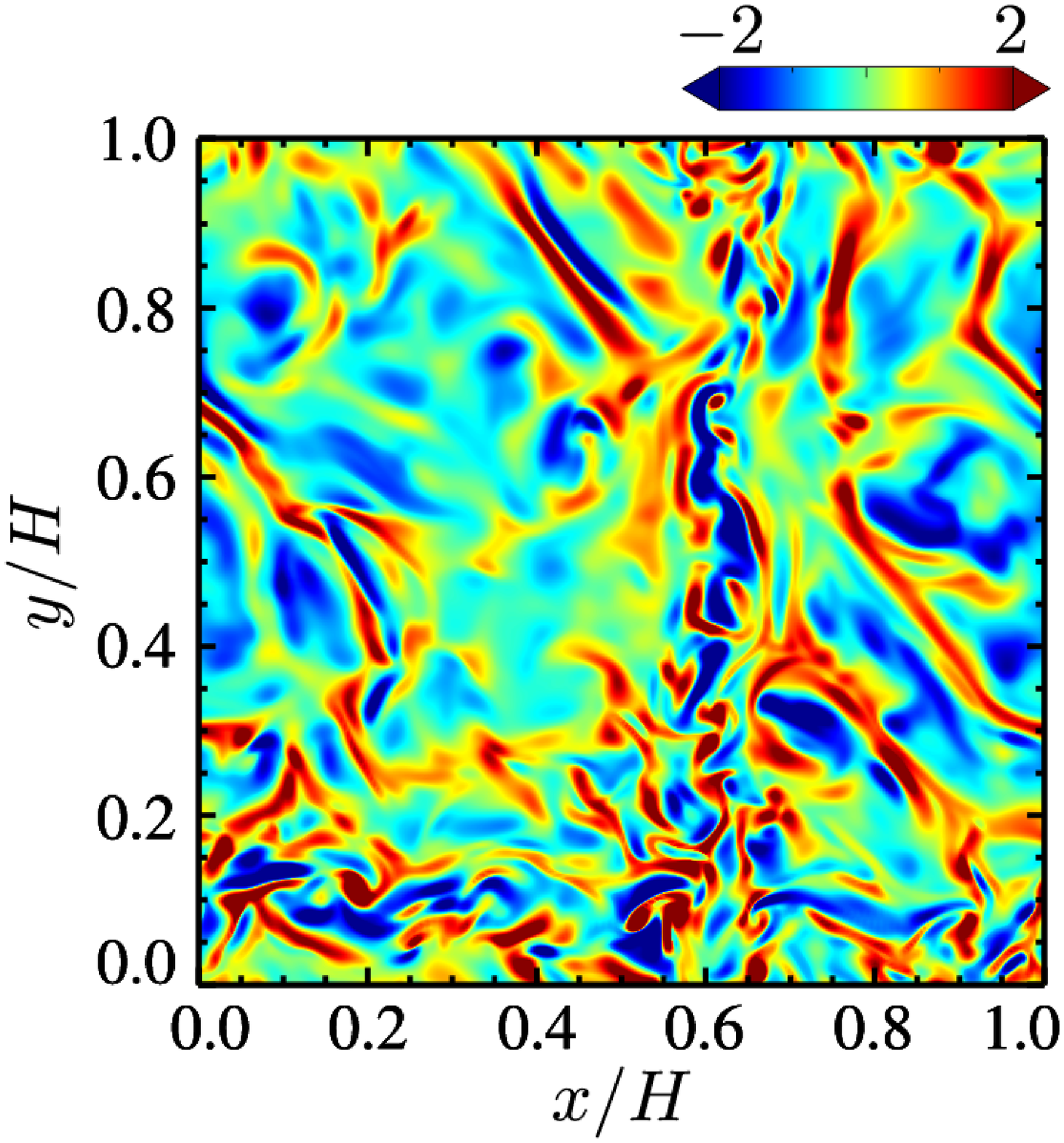}
	\end{minipage}
	\begin{minipage}{.49\linewidth}
		(\textit{b})\\
		\includegraphics[clip,width=.8\linewidth]{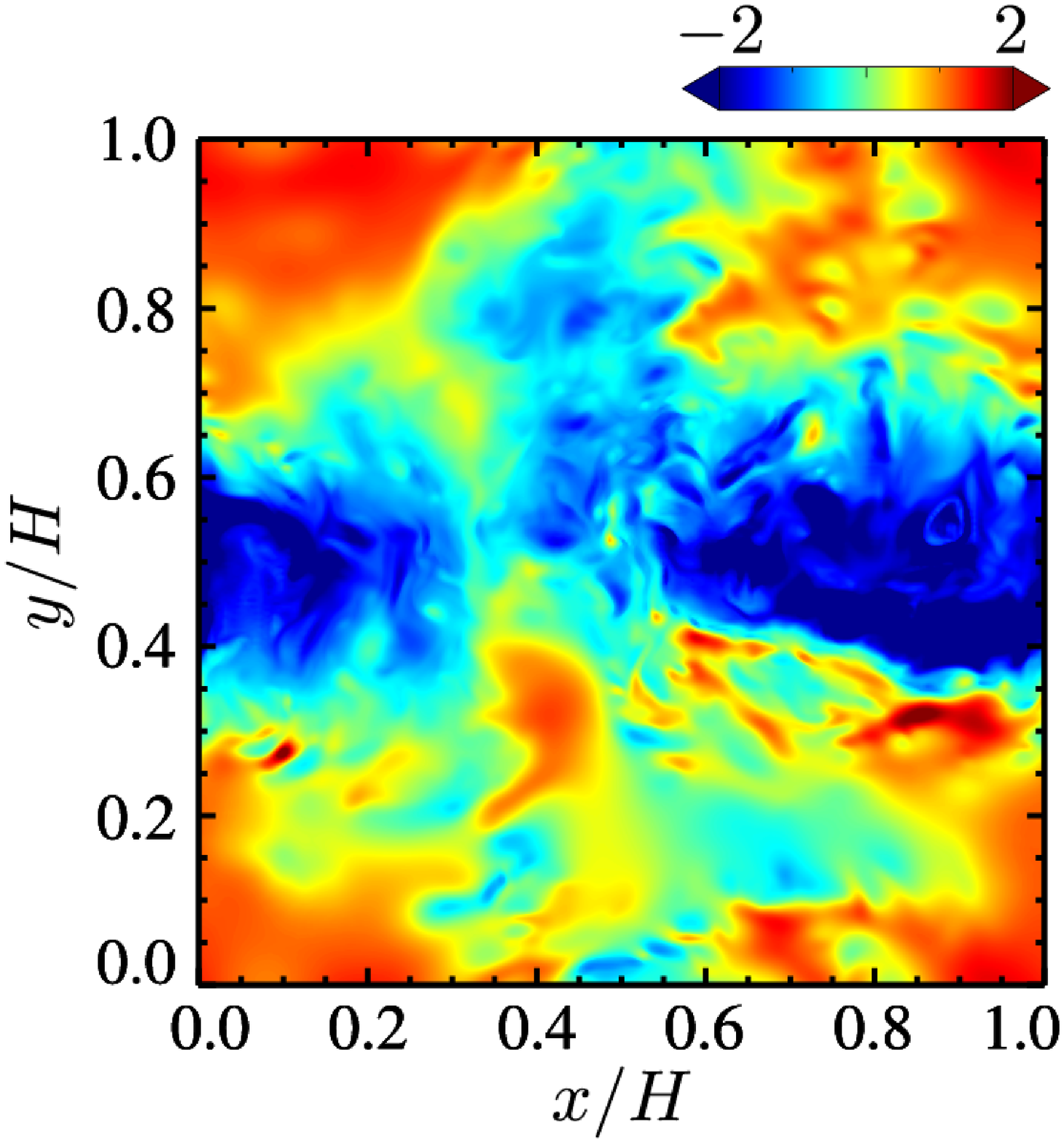}
	\end{minipage}
	
	\begin{minipage}{.49\linewidth}
		(\textit{c})\\
		\includegraphics[clip,width=.8\linewidth]{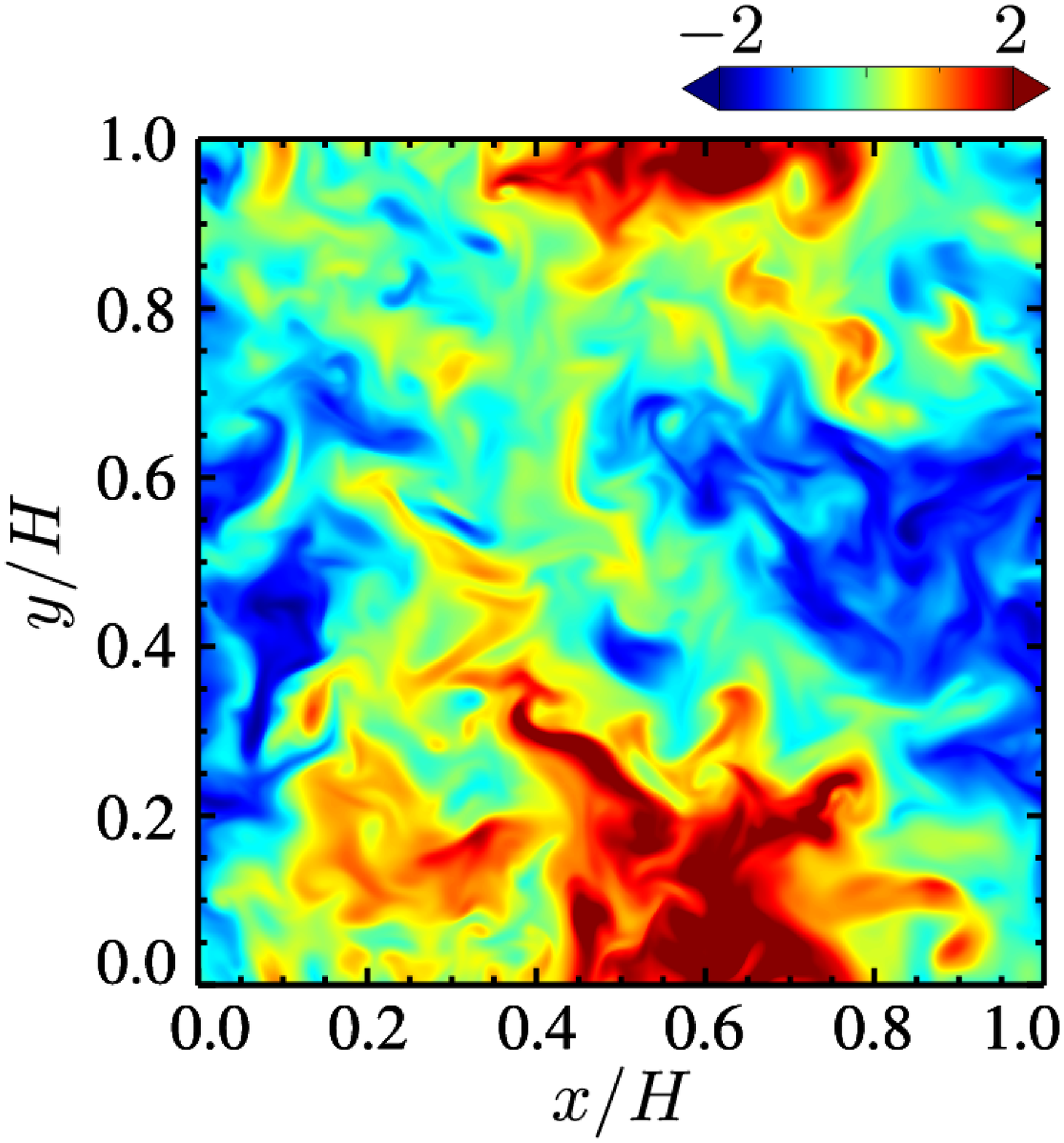}
	\end{minipage}
	\begin{minipage}{.49\linewidth}
		(\textit{d})\\
		\includegraphics[clip,width=.8\linewidth]{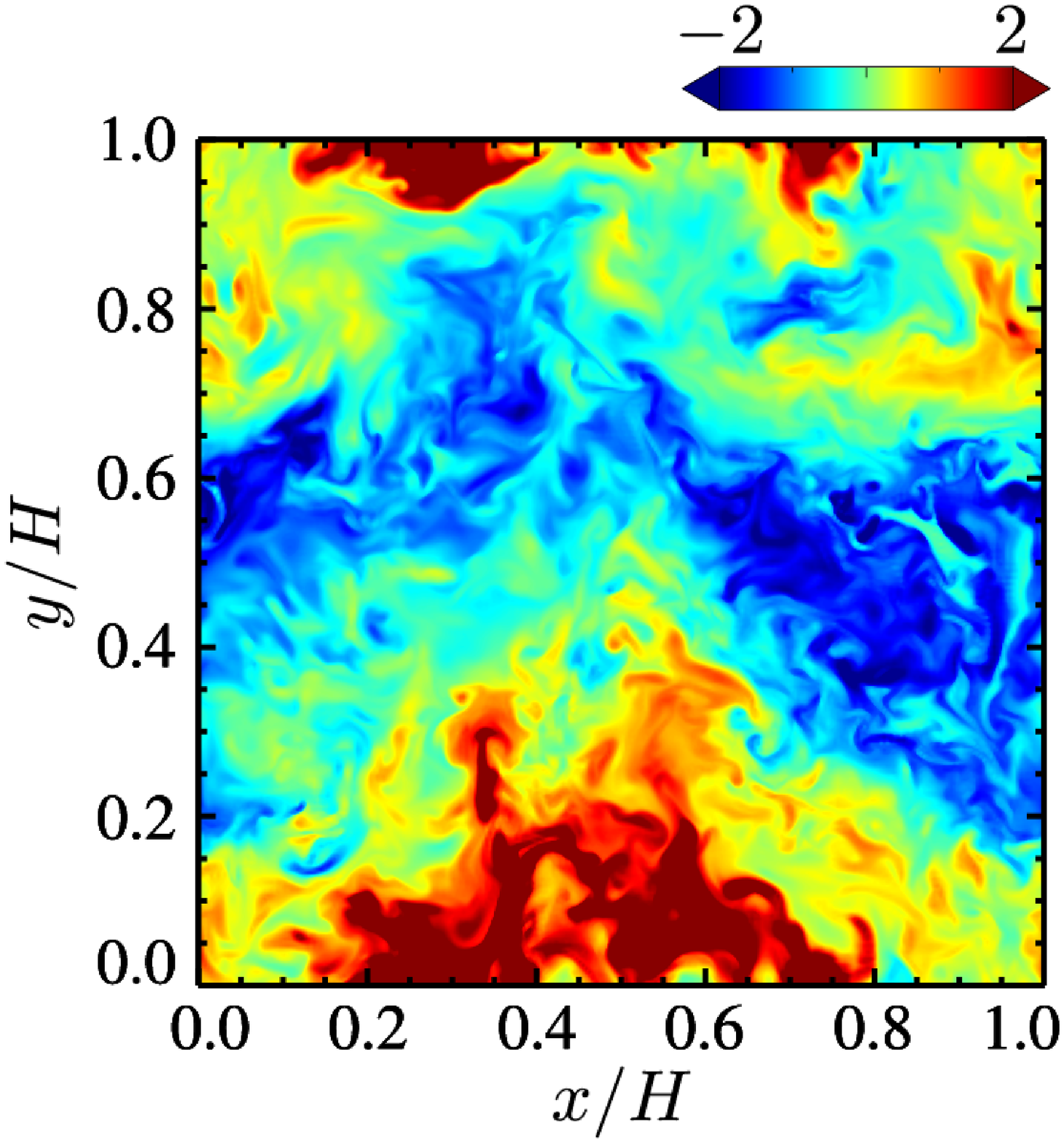}
	\end{minipage}
	\caption{Instantaneous \GK{convection} heat flux $wT$ on the near-wall plane (\textit{a,b}) \GK{$z/\delta \approx 1$ and (\textit{c,d}) the midplane $z/H=1/2$} at $Ra=10^{9}$.
		(\textit{a,c}) the impermeable case \GK{$\beta U=0$.
			(\textit{b,d}) the supercritical permeable case $\beta U=3$}.
		The heat flux \GK{$wT$ on} the horizontal plane is normalised so that its mean and \GK{standard deviation} may be \GK{zero} and \GK{unity}, respectively.
		\label{fig:uxt}}
\end{figure}

%\subsection{Turbulent thermal and flow structures}\label{sec:spatial_structure}

\GK{Let us now look into turbulence structure of thermal convection.}
Figure \ref{fig:paraview} visualises the instantaneous thermal and vortical structures in the impermeable and \GK{supercritical} permeable case at $Ra=10^{9}$.
\GK{The high-temperature thermal plumes are represented by the isotherms $T/\Delta T=0.7$, while the small-scale vortical structures are identified in terms of the positive isosurfaces of the second invariant of the velocity gradient tensor}
\begin{equation}
Q=-\frac{1}{2} \frac{\partial u_i}{\partial x_j} \frac{\partial u_j}{\partial x_i}.
\label{Q value}
\end{equation}
\GK{In the impermeable case the small-scale hot plumes are confined to the near-wall region.
In contrast, high-temperature plumes of a remarkably large horizontal length scale fully extend from the bottom wall to the top wall through the bulk in the supercritical permeable case, so that heat transfer is highly enhanced.
Recently, such promotion of large-scale circulation has been reported for the convective turbulence, which exhibits the ultimate scaling $Nu\sim Pr^{1/2}Ra^{1/2}$, in the radiatively-driven convection \citep{lepot18} and in the thermal convection between rough walls \citep{tummers19}.
Although the intensity and the size of small-scale tubular vortices playing a role in energy dissipation are different between the impermeable and the supercritical permeable cases, their spatial structure is more or less the same.}

\GK{In figure \ref{fig:uxt} are shown the snapshots of the convective heat flux $wT$ (which is proportional to local buoyancy power) on the horizontal plane in the conduction layer height $z/\delta \approx 1$ and on the midplane $z/H=1/2$.
Note that in these figures, $wT$ is normalised so that its mean and standard deviation may be \GK{zero} and \GK{unity}, respectively, in each plane of the impermeable and the supercritical permeable cases.}
The spatial distribution near the wall differs greatly between the impermeable and \GK{supercritical} permeable cases (figure \ref{fig:uxt}\textit{a,b}).
The \GK{near-wall} small-scale structures, corresponding to thermal plumes, can be observed in the impermeable case, while the large-scale structure, which is the part of the fully extended large-scale plume, appears even in the vicinity of the wall in the supercritical case.
\GK{On the midplane there is no significant difference between the impermeable and supercritical cases (figure \ref{fig:uxt}\textit{c,d}).
In the bulk region the heat transfer is dominated by large-scale convection, regardless of the difference in the near-wall dominant thermal structures.}

\begin{figure}
\centering
\begin{minipage}{.32\linewidth}
(\textit{a})\\
\includegraphics[clip,width=\linewidth]{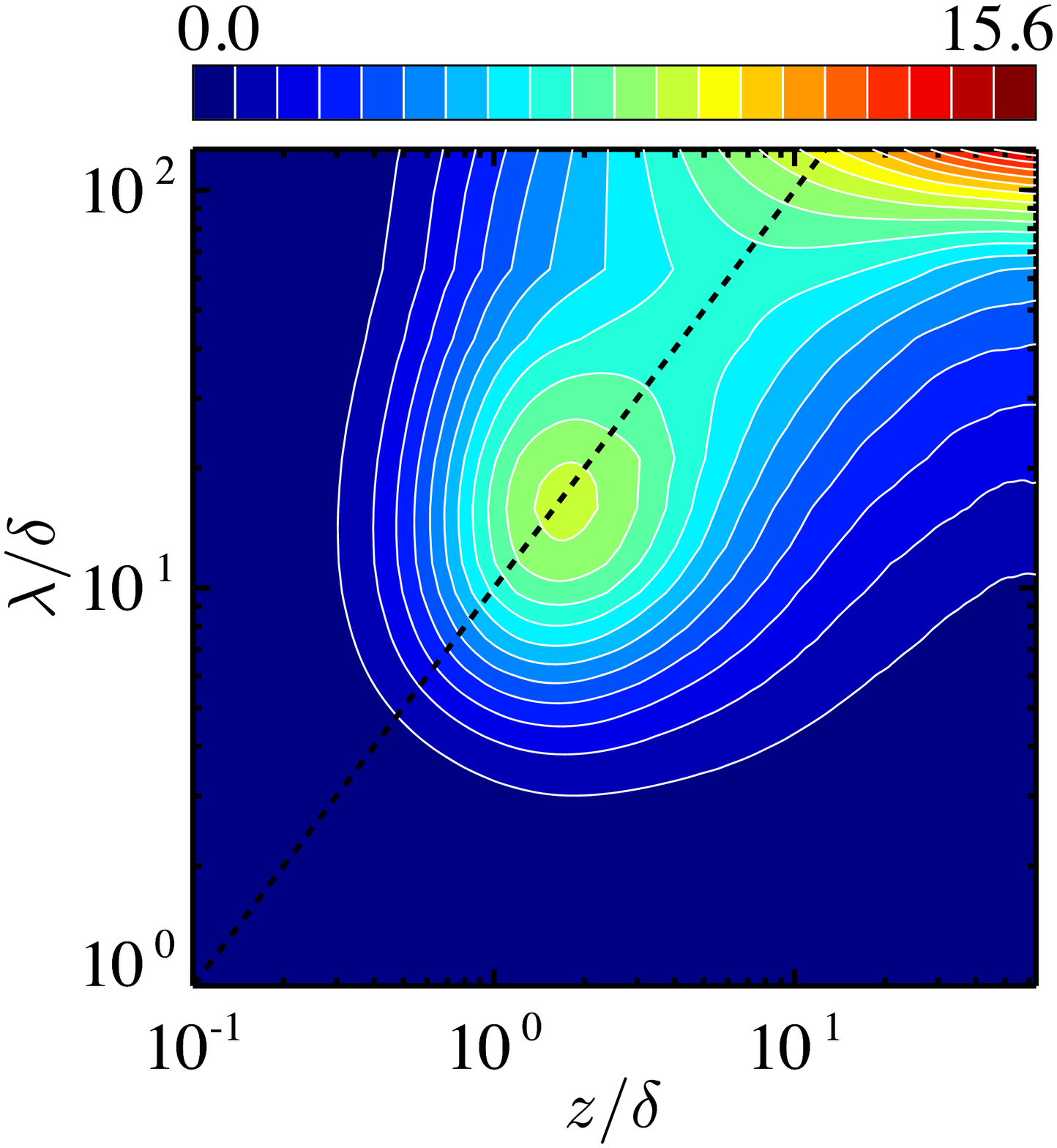}
\end{minipage}
\begin{minipage}{.32\linewidth}
(\textit{b})\\
\includegraphics[clip,width=\linewidth]{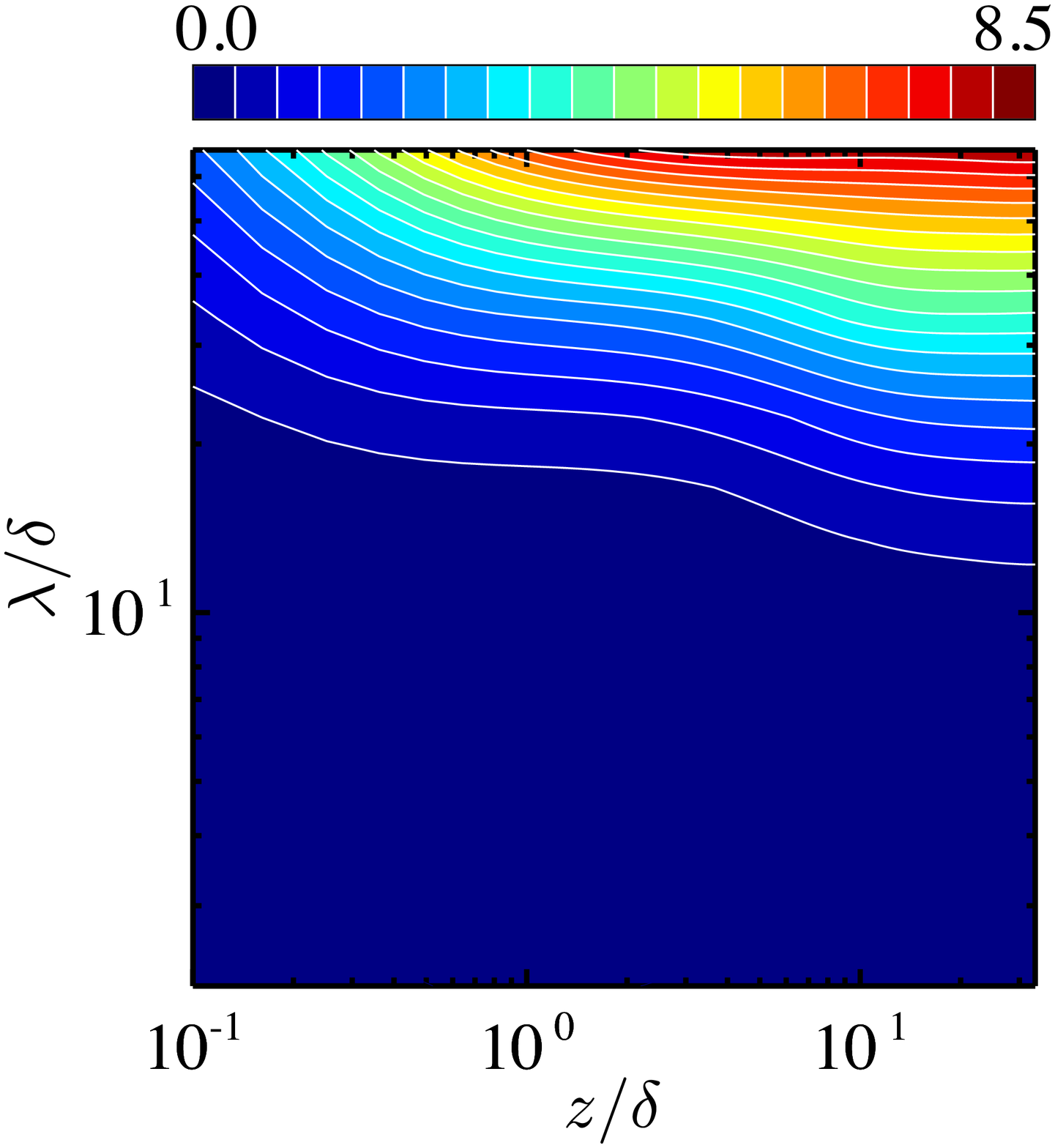}
\end{minipage}
\begin{minipage}{.32\linewidth}
(\textit{c})\\
\includegraphics[clip,width=\linewidth]{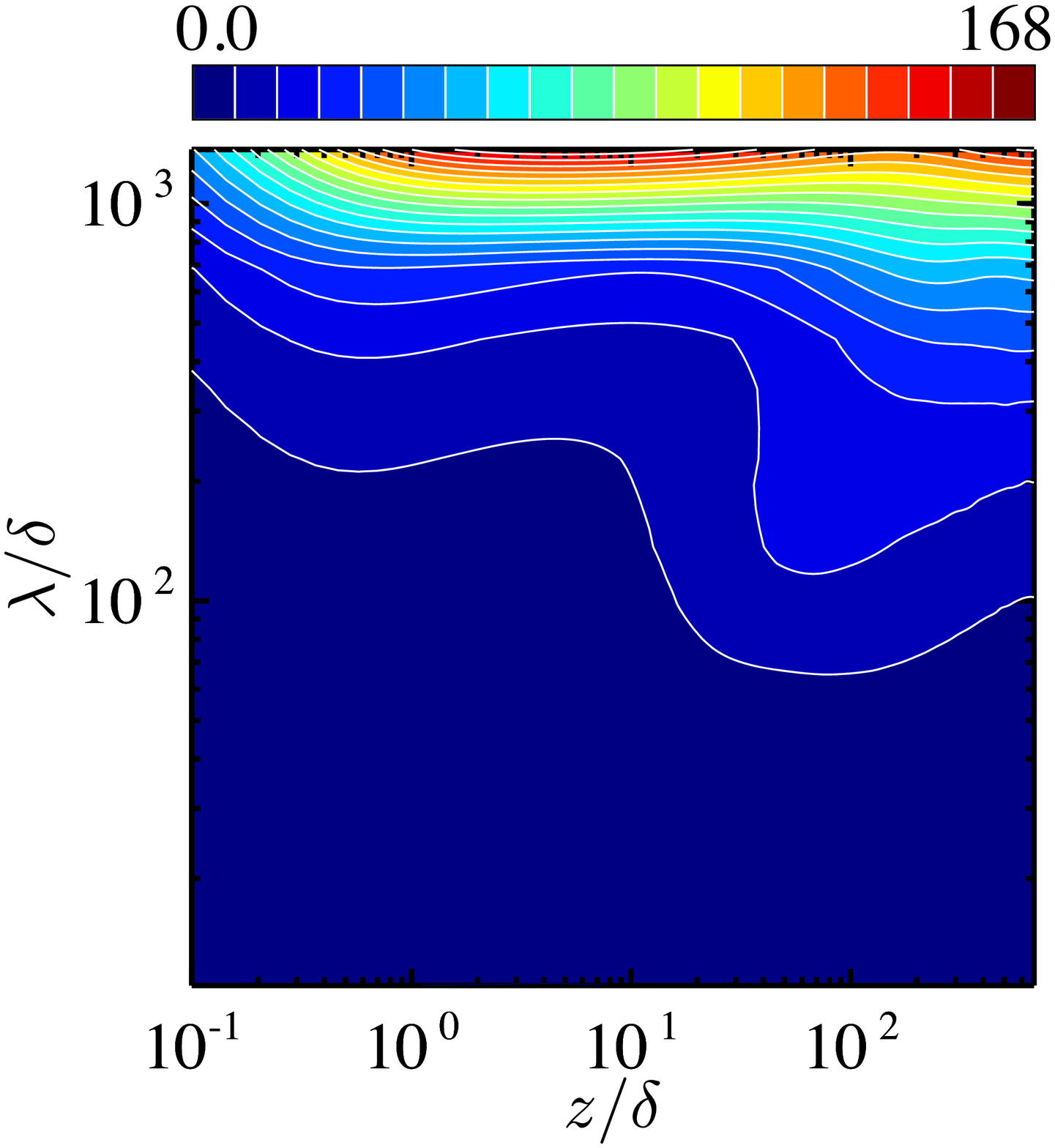}
\end{minipage}
\caption{One-dimensional premultiplied buoyancy-power spectra $k_{y}\sum_{k_{x}}\widehat{P}(k_{x},k_{y},z)$ normalised with $Ra\kappa^3/H^4$ as a function of the \GK{wavelength $\lambda=2\upi/k_y$ in the horizontal ($y$-)} direction and the distance to the bottom wall, $z$.
\GK{(\textit{a}) the impermeable case $\beta U=0$ at $Ra=10^{9}$.
(\textit{b}) the subcritical permeable case $\beta U=3$ at $Ra=10^{6}$.
(\textit{c}) the supercritical permeable case $\beta U=3$ at $Ra=10^{9}$.}
The dashed line indicates \GK{$\lambda=10z$}
\label{fig:ensp_bp}}
\end{figure}

%\subsection{Energy production by buoyancy}

In figure \ref{fig:ensp_bp} we show the one-dimensional premultiplied buoyancy-power spectra $k_{y}\sum_{k_{x}}\widehat{P}(k_{x},k_{y},z)$ as a function of the distance to the wall, $z$, and the wavelength in the \GK{horizontal ($y$-)} direction, \GK{$\lambda=2\pi/k_{y}$}.
The buoyancy-power spectra $\widehat{P}(k_x,k_y,z)$ is given by 
\begin{equation}
\widehat{P}(k_x,k_y,z)=g \alpha \GK{\mbox{Re}} \left[ \langle \widehat{w}\widehat{T}^\dagger \rangle_{t}  \right],
\end{equation}
where $\widehat{(\cdot)}$ represents the Fourier coefficients, \GK{$(k_x, k_y)$ are the wavenumbers} in the horizontal ($x$- and $y$-) directions, $\dagger$ denotes the complex conjugate and $\langle \cdot \rangle_{t}$ is the time average.
The lateral and longitudinal axes \GK{of the figures} are normalised by the conduction layer thickness \GK{$\delta$}.
$\widehat{P}$ \GK{denotes the spectrum of the energy input by buoyancy}, and \GK{it is also relevant to the spectrum of the convective heat flux} shown in figure \ref{fig:uxt}.
In the impermeable case we can see \GK{significant buoyancy power} at small scales in the vicinity of the wall, \GK{$z/\delta\sim 10^0$, leading to the near-wall thermal plumes (figure~\ref{fig:uxt}\textit{a}), in addition to greater buoyancy power corresponding to the large-scale convection in the bulk region.
The near-wall heat flux determined by the marginal instability of the thermal conduction layer gives us the scaling $Nu\sim Ra^{1/3}$ widely observed in turbulent RBC \citep{malkus54}.}
The dashed line in figure \ref{fig:ensp_bp}(\textit{a}) stands for \GK{$\lambda =10 z$}.
The spectral ridge is on this line, \GK{implying that the energy-inputted horizontal scale is proportional to the distance to the wall.}
This observation suggests that the convective heat flux \GK{exhibits} hierarchical \GK{self-similar structure} near the wall.
\GK{In the subcritical permeable case at $Ra=10^{6}$ (figure~\ref{fig:ensp_bp}\textit{b}), the Rayleigh number is too low for the small-scale plumes of $\lambda\ll L(=H)$ to appear in the near-wall region.
In the supercritical case (figure~\ref{fig:ensp_bp}\textit{c}) the spectral peak is located at the large horizontal scale $\lambda/\delta\sim 10^3$ ($\lambda/L\sim 1$) in the near-wall region $z/\delta\sim 10^0$ roughly consistent with the wall-normal position of the spectral peak of the small-scale thermal plumes, suggesting that the large-horizontal-scale plume is generated in the near-wall region by buoyancy to fully extend from there to the other wall as observed in figure~\ref{fig:paraview}({\it b}).}
\GK{This near-wall large-scale energy input corresponds to the large-scale convective heat flux shown in figure \ref{fig:uxt}(\textit{b}).}
\GK{As will be discussed in the next section \ref{sec:mechanism}, the ultimate heat transfer $Nu\sim Ra^{1/2}$} can be attributed to the generation of this \GK{long-wavelength (and so intense) thermal mode near the wall.}
In the bulk region \GK{apart} from the walls the energy is \GK{inputted at the large horizontal length scale in all the impermeable and permeable cases.
We note, however, that just in the supercritical case the energy to be inputted at the large horizontal scale in the bulk is smaller than that in the near-wall region.}

\section{Physical interpretation of scaling laws}\label{sec:mechanism}

\GK{Here we shall discuss the physical mechanisms of the ordinary scaling $Nu\sim Ra^{1/3}$ and the ultimate scaling $Nu\sim Pr^{1/2}Ra^{1/2}$ in turbulent thermal convection between impermeable and permeable walls. 
In the present study the Prandtl number has been set to unity, i.e., $Pr=1$, and thus
in this section we assume that $Pr\sim 1$ (or $\nu\sim \kappa$).

Let us start with the thermal convection in the impermeable case and the subcritical permeable case, where the temperature profile is flatter in the bulk region at higher $Ra$ and thus temperature variation is confined to the near-wall layer of the thickness of $O(\delta)$ (see figure~\ref{fig:te}{\it a,b}).
The vertical velocity is strictly zero on the impermeable walls.
In the subcritical permeable case, as shown in figure~\ref{fig:ra_wwall} and figure~\ref{fig:wrms_rbc_wall} transpiration has not been activated in the near-wall region although the walls are permeable.
In both the impermeable and subcritical cases, therefore, the near-wall vertical velocity is small in comparison to that in the bulk region.
We now suppose that in the near-wall layer with the thickness $\delta'$ of $O(\delta)$ and the temperature difference of $O(\Delta T)$ where the vertical velocity scale $U_w$ is small, the effect of viscosity is significant.
In the vertical component of the Navier--Stokes equation (\ref{ns}), the viscous term is comparable with the advection term and the buoyancy term, that is,
\begin{equation}
\nu \frac{U_w}{\delta'^2} \sim \frac{U_w^2}{\delta'} \sim  g \alpha \Delta T,
\label{eq:ns_BL}
\end{equation}
in the near-wall region.
The balance (\ref{eq:ns_BL}) between the viscous, the advection and the buoyancy terms in the equation of motion determines the near-wall velocity and the length scales as
\begin{equation}
U_w \sim Ra^{1/3}Pr^{-1/3}\nu/H \sim Ra^{-1/6}Pr^{1/6}U \sim Ra^{-1/6}U,
\label{eq:w_BL}
\end{equation}
\begin{equation}
\delta' \sim Ra^{-1/3}Pr^{1/3}H \sim Ra^{-1/3}H
\label{eq:delta_BL}
\end{equation}
(recall that $U=(g\alpha\Delta TH)^{1/2}$ and $H$ are the buoyancy-induced terminal velocity and the wall distance, respectively).
In the present DNS we have confirmed that the vertical velocity near the impermeable and subcritical permeable walls scales with $Ra^{-1/6}U$ (see figure~\ref{fig:ra_wwall}{\it a} and figure~\ref{fig:wrms_rbc_wall}).
Since the definition (\ref{eq_delta}) of the thermal conduction layer thickness implies that $\delta'\sim \delta=(H/2Nu)$, we arrive at the scaling law
\begin{equation}
Nu\sim Ra^{1/3},
\label{eq_1/3}
\end{equation}
which has been observed in RBC (i.e., the impermeable case) as well as in the subcritical permeable case (see figure\ref{fig:nu_ra}).
The scaling law $Nu\sim Ra^{1/3}$ has already been given by the several arguments on similarity \citep{priestley54}, the marginal instability \citep{malkus54} and the bulk contribution to energy and scalar dissipation \citep{GL00}.

In the bulk region of the impermeable and the subcritical cases, where the effects of viscosity or thermal conduction are no longer significant, the characteristic length scale is $H$ instead of $\delta$ (and $\delta'$), and the temperature difference with respect to the height difference of $O(H)$ and the vertical velocity scale are supposed to be $\Delta T'$ and $U_b$, respectively.
In this region the advection and the buoyancy terms balance each other out in the Navier--Stokes equation as
\begin{equation}
\frac{U_b^2}{H} \sim g \alpha \Delta T'.
\label{eq_imp_bulk}
\end{equation}
Rewriting the Nusselt number (\ref{eq_nu}) as
\begin{equation}
Nu=\frac{\langle wT \rangle_{xyt}-\kappa{\rm d} \langle T \rangle_{xyt}/{\rm d} z}{\kappa\Delta T/H}
\label{eq_heat_flux}
\end{equation}
and taking into consideration the dominance of convective heat transfer and the scaling (\ref{eq_1/3}), we have
\begin{equation}
\frac{U_b \Delta T'}{\kappa \Delta T / H} \sim Ra^{1/3}.
\label{eq_nu_wbulk}
\end{equation}
Equations (\ref{eq_imp_bulk}) and (\ref{eq_nu_wbulk}) yield the temperature difference and the velocity scale as
\begin{equation}
\Delta T' \sim Ra^{-1/9} \Delta T,
\end{equation}
\begin{equation}
U_b \sim Ra^{4/9}Pr^{-2/3}\nu/H \sim Ra^{-1/18}Pr^{-1/6} U \sim Ra^{-1/18} U.
\label{eq_wbulk}
\end{equation}
Equation (\ref{eq_wbulk}) means that the Reynolds number for thermal convection is of the order of $Ra^{4/9}Pr^{-2/3}$, being consistent with the Grossmann \& Lohse's (2000) scaling based on the energy and scalar dissipation in the bulk region.
It has been confirmed that the vertical velocity and the temperature fluctiation scale with $Ra^{-1/18}U$ and $Ra^{-1/9} \Delta T$, respectively, in the bulk region of the impermeable and subcritical permeable cases (see figure~\ref{fig:urms}{\it a,b} and figure~\ref{fig:terms}{\it a,b}).

Next we consider the thermal convection between the supercritical permeable walls.
In this case intense vertical transpiration is induced even in the vicinity of the wall in contrast to the impermeable and the subcritical cases (see figure \ref{fig:ra_wwall}).
Although the thermal conduction layer still exists on the wall, there is no near-wall layer of significant change in the vertical velocity, suggesting that the effect of the viscosity on the vertical velocity is negligible anywhere.
The vertical motion should exhibit the length scale comparable with $H$ (see figure~\ref{fig:paraview}{\it b}), and the corresponding temperature difference is of the order of $\Delta T$ even in the bulk region (recall the temperature gradient of $O(\Delta T/H)$ in figure~\ref{fig:te}{\it b} and the temperature fluctuation of $O(\Delta T)$ in figure~\ref{fig:terms}{\it f}). 
Therefore, the balance between the dominant advection and buoyancy terms in the Navier--Stokes equation (\ref{ns}) gives us
\begin{equation}
\frac{U_b^{2}}{H}\sim g\alpha\Delta T,
\label{eq:ns_permeable}
\end{equation}
leading to
\begin{equation}
U_b \sim U.
\label{eq_w_porous}
\end{equation}
The balance between the buoyancy power, the energy dissipation and the pressure power in the energy budget (\ref{total energy budget symmetry}),
\begin{equation}
\label{balance energy budget}
PrRa(Nu-1)\sim \frac{\epsilon}{\kappa^3/H^4} \sim \frac{U^3}{(\kappa/H)^3},
\end{equation}
suggests the Taylor's dissipation law (energy dissipation independent of $\nu$)
\begin{equation}
\epsilon\sim \frac{U^3}{H}
\label{taylor}
\end{equation}
and the ultimate scaling
\begin{equation}
Nu\sim Pr^{1/2}Ra^{1/2} \sim Ra^{1/2},
\label{eq_1/2}
\end{equation}
where we have taken account of $\beta U\sim 1$.
Alternatively, it follows from (\ref{eq_heat_flux}) at $z/\delta\gg 1$ and (\ref{eq_w_porous}) that
\begin{equation}
Nu\sim\frac{U_b \Delta T}{\kappa \Delta T/H} \sim Pr^{1/2}Ra^{1/2} \sim Ra^{1/2}.
\label{eq_nu_porous}
\end{equation}
The ultimate scaling $Nu\sim Ra^{1/2}$ has been suggested by \citet{kraichnan62} and by \citet{GL00} as high-$Ra$ asymptotics; however, it has not been observed in conventional RBC as yet.
In the present DNS of the supercritical permeable case, the vertical velocity has been seen to scale with $U$ in the whole region (see figure~\ref{fig:urms}{\it f}), and it has been confirmed that $Nu\sim Ra^{1/2}$ at $Ra\gtrsim 10^7$ (see figure~\ref{fig:nu_ra}).}

\begin{figure}
\centering
\begin{minipage}{.65\linewidth}
\includegraphics[clip,width=\linewidth]{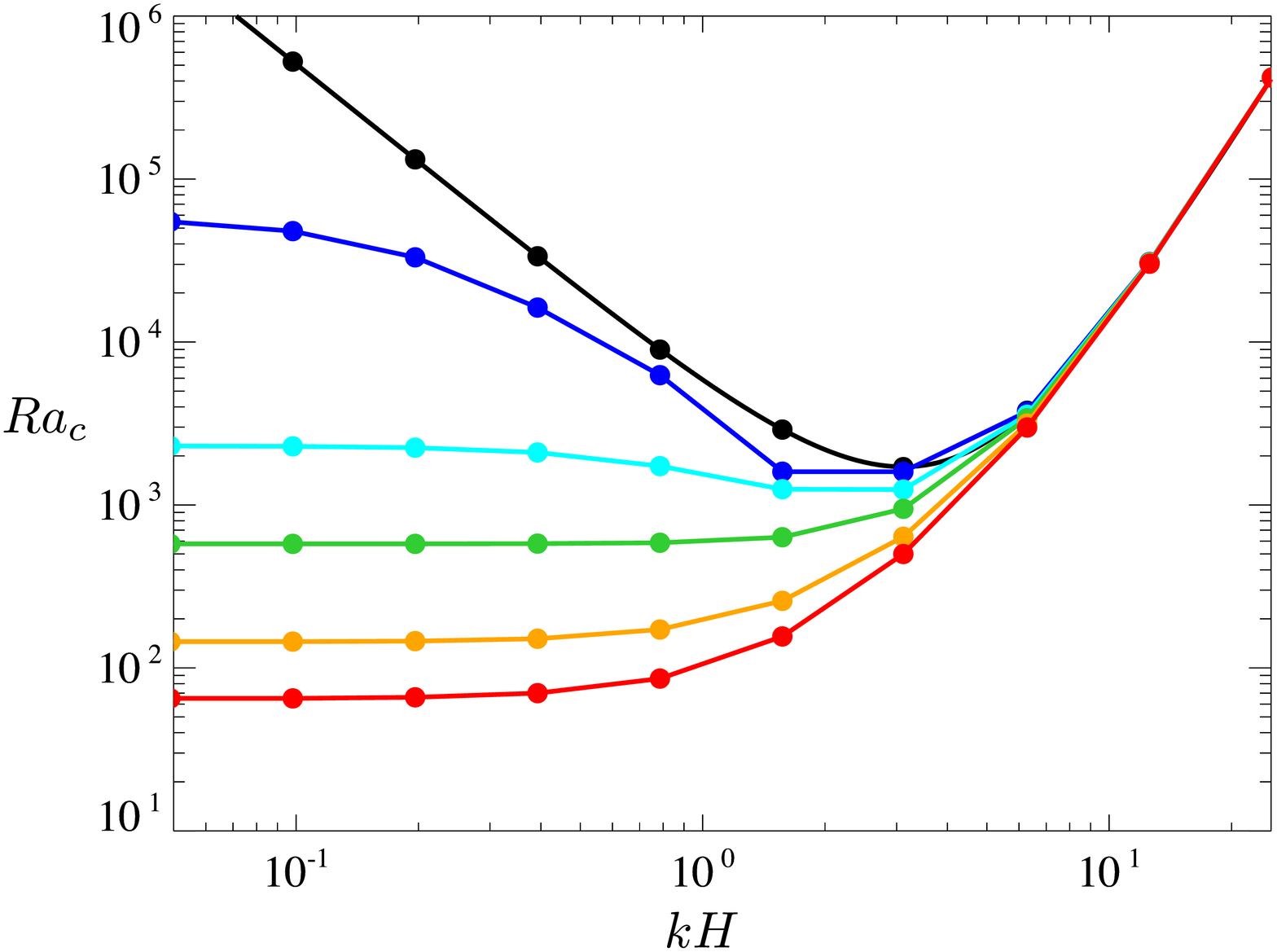}
\end{minipage}
\caption{\GK{The critical Rayleigh number $Ra_c$ of the onset of two-dimensional thermal convection between impermeable and permeable walls as a function of the horizontal wavenumber $k$.
The black symbols denote the impermeable case $\beta U=0$.
The other lines with the symbols represent the permeable case: blue, $\beta U=0.1$; cyan, $\beta U=0.5$; green, $\beta U=1$; orange, $\beta U=2$; red, $\beta U=3$.
Black curve stands for the analytical marginal stability relation given by \citet{prosperetti11} for RBC (i.e. the impermeable case).}
\label{fig:stability}}
\end{figure}

\GK{In the above discussions we have considered the difference in the vertical length scale of thermal convection, $\delta$ and $H$, in the impermeable (and subcritical permeable) case and the supercritical permeable case and its crucial consequences on the scaling properties of heat transfer.
The key to the difference in the vertical length scale is the excitation of transpiration in the near-wall region of the permeable wall.
As suggested in figure~\ref{fig:ensp_bp}, there should be different convection modes of the instabilities in a thermal conduction layer, one of which is the small-scale thermal plume in the impermeable (and subcritical permeable) case, and the other of which is the large-scale plume extending to the other wall in the supercritical permeable case.
The excitation of the near-wall transpiration velocity on the permeable wall could be attributed to the different length of convection instability from that on the impermeable wall.
In order to identify the different length of the instability, we have performed the linear stability analysis of a conduction state between the impermeable and permeable walls by conducting DNS in conjunction with the Arnoldi iteration.
Although the onset of thermal convection in the conduction state is distinct from that in the thermal conduction layer of turbulent convection, we could expect their qualitative similarity.
	
Figure~\ref{fig:stability} presents the onset Rayleigh number of thermal convection between impermeable and permeable walls.
In the impermeable case we confirm the known lowest value $Ra_c=1708$ for $kH=3.117$ (black symbols).
In the permeable case, on the other hand, much larger-scale thermal convection (for much smaller $kH$) can arise from the instability (colour lines with symbols).
If such larger-horizontal-scale thermal plume appears in the thermal conduction layer of convective turbulence, then the plume should also possess a larger  vertical length scale to induce the significant vertical velocity.
Actually the large-horizontal-scale thermal plume has been observed to extend from the near-wall region to the other wall in turbulent convection on the supercritical permeable walls (see figure~\ref{fig:paraview}{\it b} and figure~\ref{fig:uxt}{\it b}).
The critical transition to the ultimate heat transfer observed in figure~\ref{fig:nu_ra} and figure~\ref{fig:ra_wwall} could be a consequence of the exchange of near-wall unstable convection modes on the permeable wall.
	
In DNS of the permeable case we have increased the horizontal period $L$ in the range of $1\leq L/H\leq 4$ and have observed stronger convection in a wider periodic box of larger $L/H$ for $\beta U=3$.
This would be because longer-wavelength convection is more significant as a result of the convection instability (see figure~\ref{fig:stability}).
Smaller $\beta U\sim 10^{-1}$ would, however, lead to an optimal length scale of convection and thus no significant dependence of convective turbulence on the horizontal domain size.}

\section{Summary and outlook}\label{sec:summary}

We have performed the three-dimensional \GK{direct numerical simulation} (DNS) of \GK{turbulent} thermal convection between horizontal no-slip, permeable walls \GK{with a distance $H$ and a constant temperature difference $\Delta T$}.
\GK{On the no-slip wall surfaces $z=0$, $H$ the vertical transpiration velocity has been assumed to be proportional to the local pressure fluctuation \citep{jimenez01}, i.e. $w=-\beta p'/\rho$, $+\beta p'/\rho$ mimicking a Darcy-type permeable wall \citep[pp. 223--224]{batchelor67}.}
A zero net mass flux through the permeable wall is instantaneously ensured, and convective turbulence is driven \GK{only} by buoyancy without \GK{any additional energy inputs.}
\GK{The permeability parameter is set to $\beta U=0$ (an impermeable case) and  $\beta U=3$ (a permeable case) where $U=(g\alpha\Delta TH)^{1/2}$ is} the buoyancy-induced terminal velocity.
DNS has been carried out at the Rayleigh number up to $Ra=10^{11}$ in the impermeable case and $Ra=10^{10}$ in the permeable case for fixed Prandtl number $Pr=1$.
We have found that the wall permeability leads to \GK{the critical transition of the Nusselt number scaling with the Rayleigh number from $Nu\sim Ra^{1/3}$ to the ultimate scaling $Nu\sim Ra^{1/2}$ as $Ra$ increases.}

In the subcritical regime $10^{6}\lesssim Ra\lesssim10^{7}$ \GK{we have found the scaling law $Nu\sim Ra^{1/3}$ commonly observed in turbulent Rayleigh--B\'enard convection (RBC) although on the permeable wall, there are weak vertical velocity fluctuations of the order of $Ra^{-1/6}U$ comparable with the velocity scale of near-wall small-scale thermal plumes in RBC (i.e. the impermeable case).
The mean temperature gradient becomes small in the bulk region as $Ra$ increases, and temperature fluctuations scale with $Ra^{-1/9}\Delta T$ in the bulk.}

In the supercritical regime $10^{7}\lesssim Ra\lesssim10^{10}$, on the other hand, the ultimate scaling $Nu\sim Ra^{1/2}$ \GK{has been found}.
\GK{In this supercritical regime the mean temperature profile exhibits a steeper gradient in the very-near-wall thermal conduction layers at higher $Ra$ while a finite value of the temperature gradient remains in the bulk region, implying temperature fluctuations of $O(\Delta T)$, in contrast to the vanishing bulk temperature gradient in the impermeable and subcritical permeable cases.
This situation is very different from convective turbulence without horizontal walls \citep{calzavarini05,pawar16}, in which there is no thermal conduction layer and the ultimate scaling has also been observed.}
\GK{In the supercritical case the significant transpiration velocity is induced even in the vicinity of the wall.
The vertical velocity fluctuation scales with $U$ at any height.
Although the vertical velocity fluctuation is suppressed near the permeable wall in comparison to the bulk region, there is no near-wall layer of large change in the vertical velocity, suggesting that the effect of viscosity is negligible even in the near-wall region. 
In such `wall-bounded' convective turbulence the vertical fluid motion exhibits the large length scale of $O(H)$ in the whole region, and the buoyancy acceleration by the temperature difference of $O(\Delta T)$ can achieve the vertical velocity comparable with the terminal velocity $U$.
The ultimate heat transfer is attributed to the resulting large-scale strong plumes extending from the near-wall region of one permeable wall to the other wall.
The balance between buoyancy power, energy dissipation and pressure power on the permeable walls in the total energy budget equation provides us with the Taylor's dissipation law $\epsilon\sim U^3/H$ as well as the ultimate scaling $Nu\sim Ra^{1/2}$.
The key to the achievement of the ultimate heat transfer is the activation of transpiration in the near-wall region of the permeable wall, leading to the large-scale and so intense vertical fluid motion.
The excitation of transpiration is considered to be a consequence of near-wall larger-horizontal-scale unstable convection mode on the permeable wall, distinct from that on the impermeable or less permeable wall.}

Finally, we \GK{would like to suggest the possibility of the ultimate heat transfer in physical experiments.}
The properties of the present permeable wall can be estimated as a \GK{porous wall of many fine through holes in the vertical direction with a constant-pressure plenum chamber underneath (or overhead).
We install so many holes in the wall that the entire surface of the wall is almost covered by the holes.  
Supposing the flow through the holes to be laminar and thus be represented by the Hagen--Poiseuille flow, we have its mean velocity}
\begin{equation}
\overline{w}=\frac{d^2}{32\nu l}\frac{\Delta p}{\rho},
\label{eq:exp1}
\end{equation}
where $d$, $l$ and $\Delta p$ represent the diameter of the \GK{holes}, the thickness of the \GK{wall} and the pressure drop through \GK{the wall (or the pressure difference with respect to the constant pressure in the plenum chamber)}, respectively.
From the permeable boundary condition (\ref{eq_darcy}), the permeability coefficient $\beta$ can be expressed \GK{rigorously} as 
\begin{equation}
\beta=\frac{d^{2}}{32\nu l},
\label{eq:exp2}
\end{equation}
and its dimensionless expression is
\begin{equation}
\displaystyle
\beta U=\frac{1}{32}{\left( \frac{d}{H} \right)}^{2} \frac{H}{l}Pr^{-1/2}Ra^{1/2}.
\label{eq:exp3}
\end{equation}
If all the pressure power on the permeable wall \GK{by thermal convection} (being an energy sink \GK{for the convection}) is consumed to drive the flow in the porous \GK{wall}, \GK{their energy balance,}
\begin{equation}
\left. \frac{\beta}{\rho^{2}H} {\left< p^{2} \right>}_{xyt} \right|_{\rm wall}
\GK{\equiv}\left. \frac{1}{\beta H}{\left< w^{2} \right>}_{xyt} \right|_{\rm wall}
\sim\overline{w}\frac{\Delta p}{\rho \GK{l}}
=\frac{1}{\beta \GK{l}}\overline{w}^{2},
\label{eq:exp4}
\end{equation}
\GK{suggests that
\begin{equation}
l/H\sim1,
\label{eq:exp5-1}
\end{equation}
where we have used the reasonable relation $\left< w^{2} \right>_{xyt}\big |_{\rm wall}\sim \overline{w}^{2}$.
It follows from (\ref{eq:exp3}) that
\begin{equation}
d/H\sim (\beta U)^{1/2} Pr^{1/4}Ra^{-1/4}.
\label{eq:exp5-2}
\end{equation}

Now we map the above estimates onto turbulent thermal convection between the permeable walls.}
In the supercritical permeable case \GK{$\beta U=3$ at $Ra\sim 10^{9}$ for $Pr\sim 1$, equation (\ref{eq:exp5-2}) tells us that $d/H\sim10^{-2}$.
Since the RMS vertical velocity on the supercritical permeable wall is approximately $10\%$ of $U$ (see figure \ref{fig:ra_wwall}\textit{b}),
the Reynolds number of the flow in the holes} might be $Re=\overline{w}d/\nu\sim10^{1}$, implying that the flow is laminar.
Therefore, we \GK{may say that the properties of the supercritical permeable walls can be implemented by using the above porous walls to achieve the ultimate heat transfer in physical experiments.}

\begin{figure}
\centering
\begin{minipage}{.8\linewidth}
\includegraphics[clip,width=\linewidth]{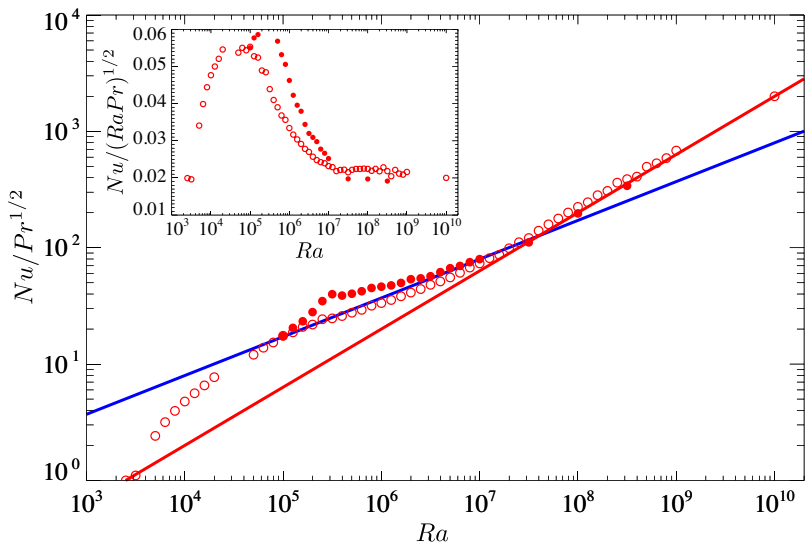}
\end{minipage}	
\caption{\GK{The Nusselt number} $Nu$ compensated by $Pr^{1/2}$ as a function of \GK{the Rayleigh number} $Ra$.
The filled and open circles respectively represent the present DNS data for $Pr=7$ and $Pr=1$ in the permeable case \GK{$\beta U=3$}.
The red and blue line indicate \GK{$Nu\sim Pr^{1/2}Ra^{1/2}$} and \GK{$Nu\sim Ra^{1/3}$}, respectively.
The inset shows $Nu$ compensated by $Pr^{1/2}Ra^{1/2}$.
\label{fig:nu_ra_pr7}}
\end{figure}

\begin{figure}
\hspace*{.05\linewidth}
%\centering
\begin{minipage}{.49\linewidth}
(\textit{a})\\
\includegraphics[clip,width=0.8\linewidth]{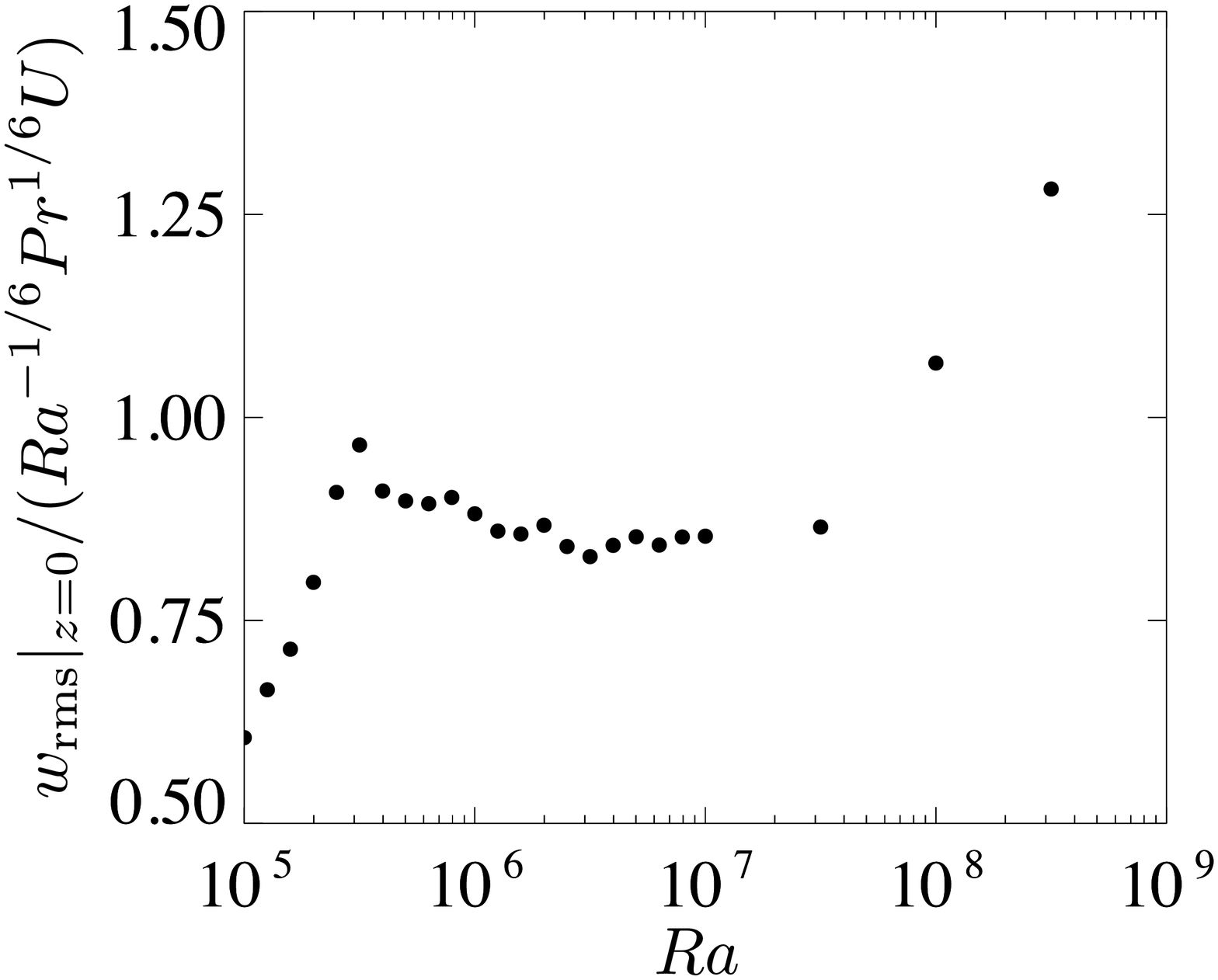}
\end{minipage}	
\begin{minipage}{.49\linewidth}
(\textit{b})\\
\includegraphics[clip,width=.8\linewidth]{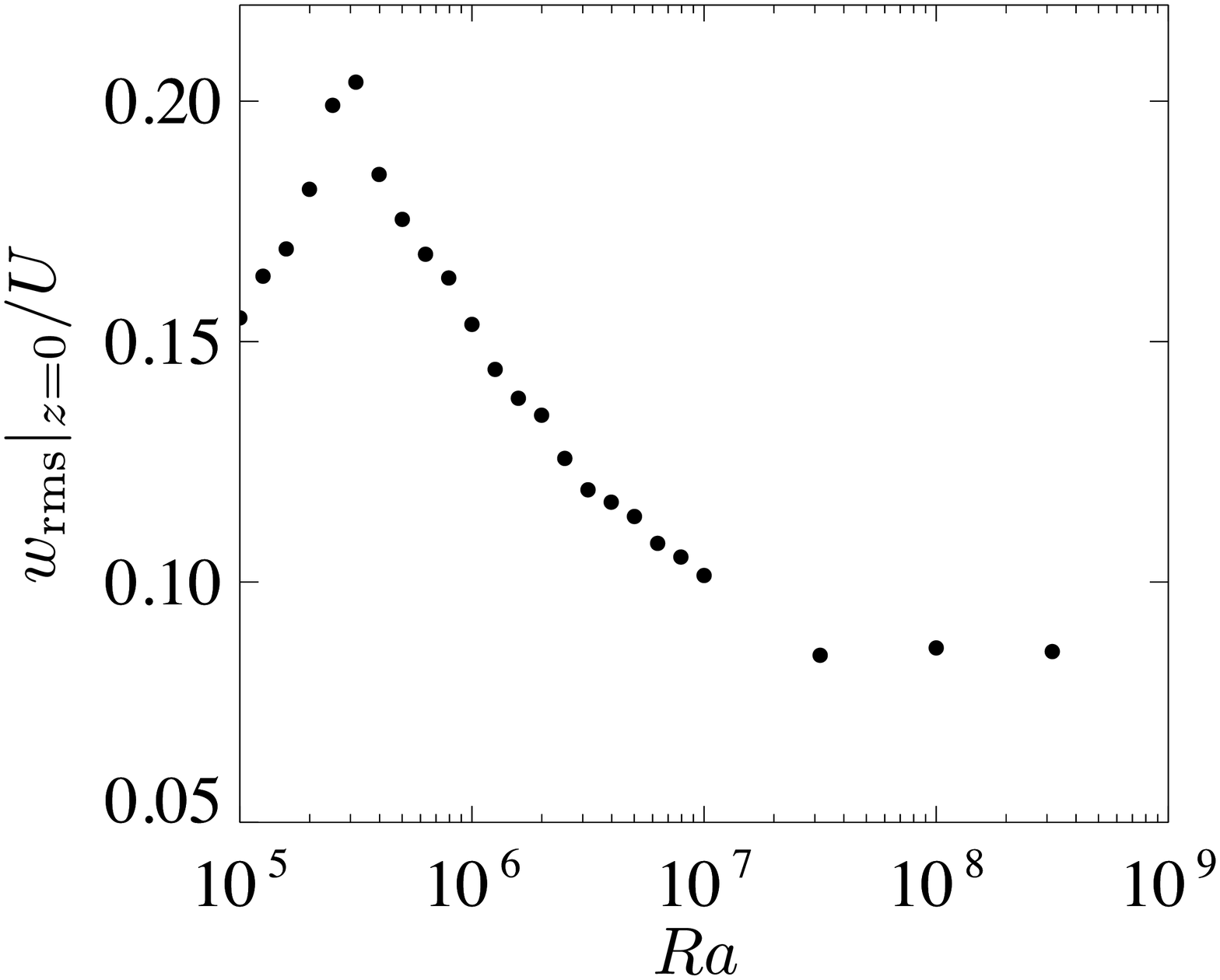}
\end{minipage}
\caption{RMS vertical velocity \GK{on} the wall $z=0$ \GK{normalised} by (\textit{a}) \GK{$Ra^{-1/6}Pr^{1/6}U$} and (\textit{b}) \GK{$U$} in the permeable case \GK{$\beta U=3$} for $Pr=7$.
\label{fig:ra_wwall_pr7}}
\end{figure}

\appendix

\section{Prandtl-number dependence}\label{sec:prandtl}

\GK{In order to examine the effects of the Prandtl number $Pr$ on the scaling of the Nusselt number $Nu$ with the Rayleigh number, we have performed DNS of turbulent convection between the permeable walls for $Pr=7$. 
We inspect the ultimate scaling law}
\begin{equation}
Nu\sim Pr^{1/2}Ra^{1/2}.
\label{eq:nu_pr7}
\end{equation}
Figure \ref{fig:nu_ra_pr7} shows $Nu$ compensated by $Pr^{1/2}$ as a function of the Rayleigh number $Ra$ at $Pr=7$ and $Pr=1$ \GK{for the horizontal period $L/H=1$} and the permeability \GK{$\beta U=3$}.
The transition to the ultimate scaling $Nu\sim Pr^{1/2}Ra^{1/2}$ is also observed for $Pr=7$.
In the supercritical $Ra$ range, the compensated $Nu$-plots roughly collapse on a single line.
The scaling behaviour in the subcritical $Ra$-range for $Pr=7$ is different from that for $Pr=1$, and the transition point seems to have a slight $Pr$ dependence.

The critical transition is also observed in the RMS vertical velocity on the wall for $Pr=7$ as shown in figure \ref{fig:ra_wwall_pr7}.
We have confirmed that the other turbulent statistics and structures are similar to those observed for $Pr=1$.

\bibliography{ref}
\bibliographystyle{jfm}

\end{document}